\RequirePackage{fix-cm}
\documentclass[smallextended]{svjour3}       
\smartqed  

\usepackage{algorithmic}
\usepackage{graphicx}
\usepackage{textcomp}
\usepackage[dvipsnames,table]{xcolor}
\usepackage{xspace}
\usepackage{fontawesome}
\usepackage{multirow}
\usepackage{rotating}
\usepackage{tikz}
\usepackage{soul}
\usepackage{makecell}
\usepackage{url}
\usepackage{enumitem}
\usepackage{booktabs}
\usepackage{fontawesome}
\usepackage{pifont}
\usepackage{tcolorbox}
\usepackage[caption=false]{subfig}
\usepackage[export]{adjustbox}
\usepackage{amsmath,amsfonts}
\usepackage{MnSymbol}
\usepackage{balance}
\usepackage[authoryear]{natbib}
\usepackage[colorlinks=true,linkcolor=blue,citecolor=blue,urlcolor=blue]{hyperref}

\newcommand{\ie}{\emph{i.e.,}\xspace}
\newcommand{\eg}{\emph{e.g.,}\xspace}
\newcommand{\etc}{etc.\xspace}

\newcommand{\secref}[1]{Section~\ref{#1}\xspace}

\newcommand{\figref}[1]{Fig.~\ref{#1}\xspace}
\newcommand{\tabref}[1]{Table~\ref{#1}\xspace}

\newcommand{\cg}{ChatGPT\xspace}
\newcommand{\cp}{Copilot\xspace}

\newcommand{\manually}{1,501\xspace}

\newcommand{\categoriesAll}{52\xspace}

\newcommand{\totalNewlyInspected}{1,161\xspace}
\newcommand{\newlyInspectedChatGPT}{728\xspace}
\newcommand{\newlyValidChatGPT}{343\xspace}
\newcommand{\newlyInspectedCopilot}{933\xspace}
\newcommand{\newlyValidCopilot}{329\xspace}
\newcommand{\finalCopilot}{281\xspace}
\newcommand{\newCategoriesAllChatGPT}{61\xspace}
\newcommand{\totalNewValid}{624\xspace}
\newcommand{\categoriesCopilot}{39\xspace}

\newcommand{\rev}[1]{\textcolor{black}{#1}}

\newboolean{showcomments}

\setboolean{showcomments}{true}

\ifthenelse{\boolean{showcomments}}
{\newcommand{\nb}[2]{
		\fbox{\bfseries\sffamily\scriptsize#1}
		{\sf\small$\blacktriangleright$\textit{#2}$\blacktriangleleft$}
	}
	
}
{\newcommand{\nb}[2]{}
	
}

\begin{document}

\title{Developers and Generative AI: A Study of Self-Admitted Usage in Open Source Projects
}

\titlerunning{A Study of Self-Admitted Usage in Open Source Projects}        

\author{Rosalia Tufano         \and
        Federica Pepe \and
        \\Fiorella Zampetti \and
        \\Antonio Mastropaolo          \and
        Ozren Dabi\'c                  \and
        Massimiliano Di Penta        \and
        Gabriele Bavota
}


\institute{R. Tufano and G. Bavota \at
              SEART @ Software Institute \\
              Universit\`a della Svizzera italiana, Switzerland \\
              \email{\{rosalia.tufano, gabriele.bavota\}@usi.ch}          
        \and
           F. Pepe and F. Zampetti and M. Di Penta \at
           University of Sannio, Italy \\
           \email{f.pepe8@studenti.unisannio.it, \{dipenta, fzampetti\}@unisannio.it}  
           \and
           A. Mastropaolo \at
           William \& Mary, United States \\
           \email{amastropaolo@wm.edu}
          \and
           O. Dabi\'c \at
           JetBrains Research, Serbia \\
           \email{dabic.ozren@gmail.com}
}

\date{Received: date / Accepted: date}

\maketitle

\begin{abstract}
The availability of generative Artificial Intelligence (AI) tools such as ChatGPT or GitHub Copilot is reshaping the way in which software is developed, evolved, and maintained. Oftentimes, developers leave traces of such an usage in software artifacts. This allows not only to understand how AI is used in software development, but also to let others be aware how such software artifacts were created, e.g., for licensing or trustworthiness purposes.
This paper---building upon our preliminary work presented at MSR 2024---aims at qualitatively investigating on the self-admitted use of two very popular generative AI tools---ChatGPT and GitHub Copilot---in software development. 
To this aim, we mined GitHub for such traces, by looking at commits, issues and pull requests (PRs). Then, through a manual coding, we create a taxonomy of 64 different ChatGPT and GitHub Copilot usage tasks, grouped into 7 categories.
By repeating our previous analysis two years after and by extending it to GitHub Copilot, we show how the usage avenues have been expanded, the extent to which developers perceived such a generative AI usage useful, and whether some concerns occurring more than one year ago are no longer present.
The taxonomy of tasks we derived from such a qualitative study provided (i) developers with valuable insights into how generative AI can be integrated into their workflows, and (ii) researchers with a clear overview of tasks that developers perceive as well-suited for automation


\end{abstract}

\sloppy

\section{Introduction} \label{sec:intro}

Software engineering (SE) researchers proposed several solutions aimed at automating non-trivial software-related tasks. At first, classification tasks such as fault prediction, traceability recovery, and reviewer recommendation were tackled. With the surge of Artificial Intelligence (AI) in SE, researchers shifted their attention towards the more complex generative tasks, such as code generation \citep{svyatkovskiy:fse2020,ciniselli:tse2021}, code review \citep{tufano:icse2021,tufano:icse2022,li:esecfse2022,thongtanunam:icse2022}, and bug-fixing \citep{tufano:ase2018,berabi:icml2021,sobania2023analysis}.  Most of these works relied on deep learning (DL) models explicitly trained for the target task. For instance, to train a DL model for automated bug fixing, one can extract bug-fixing commits from software repositories to construct a dataset comprising pairs of $<$\textit{buggy, fixed}$>$ code snippets—where the ``\textit{buggy}'' version represents the code before the fix, and the ``\textit{fixed}'' version corresponds to the corrected (and ideally bug-free) code.

The advent of Large Language Models (LLMs) further changed the landscape of automation in SE: LLM-based  chat bots such as \cg \citep{chatgpt} or tools within Integrated Development Environments (IDEs) such as GitHub Copilot \citep{copilot} (in the following referred to as  ``Copilot") allow developers to receive help for a wide number of tasks, including, but not limited to, designing an architecture, prototype a feature in a given programming language, and generate tests for existing code~\citep{HouZLYWLLLGW24}.

Despite growing interest in the capabilities of these tools when applied to software-related tasks, there is still limited empirical understanding of how they are actually used in practice, if not studies based on developers' perception \citep{abs-2406-07765} which indicated how generative AI is typically used for boring and repetitive tasks.
 
Also, the increasing usage of generative AI in software development poses the need to properly keep track of such an usage, by ``self-admitting'' whether---and, possibly, the extent to which---a given software development task was supported by generative AI, and whether such a support turned out to be useful. \rev{In some circumstances, the self-admission is even automatically generated, \eg when creating a pull request (PR) or generating a commit message with tools such as GitHub Copilot or Claude-code \citep{anthropic2025claude_code_best_practices}.}
	
\rev{On the one hand, the presence of generative AI usage self-admission serves as a traceability mechanism to raise awareness on the provenance of certain software artifacts or understand the rationale of certain decisions, with the ultimate goal of assessing their quality, trustworthiness, and licensing/copyright compliance. Moreover, establishing such a provenance could even serve to augment Software Bills of Materials (SBOMs) with information about AI-generated artifacts.}

\rev{On the other hand, the availability of generative AI usage self-admission related to different types of software engineering tasks may be used to accumulate knowledge about generative AI advantages and limitations in software development. This is very similar to what previous research has done by studying the different types of self-admitted technical debt (SATD) \citep{PotdarS14,BavotaR16,CasseeZNSP22}. While these taxonomies did not necessarily cover the whole technical debt spectrum, they provide a good overview of different technical debt types how developers admit them. On the same line, we conjecture that generative AI usage self-admission helps to provide an overview of its applications, pros and cons.}

In our MSR 2024 paper \citep{tufano:msr2024}, we partially filled this gap via a mining-based study. In particular, we mined all commits, issues, and pull requests (PRs) from GitHub that match the keyword ``ChatGPT". Then, we extracted n-grams surrounding the word \cg and manually reviewed them for further filtering. This allowed us to filter the initial sample to reduce the chance of false positives, \eg an issue mentioning \cg but not using it for the automation of a task. Then, we performed an open coding based on card sorting \citep{spencer2009card} on all \manually candidate instances we identified, classifying the \cg purpose for each instance or discarding the instance as a false positive.  As a result, we obtained a taxonomy of purposes for \rev{self-admitting} \cg \rev{usage} in the automation of a software-related task. The taxonomy features seven root categories and \categoriesAll categories in total. A limitation of our initial study is that it focused exclusively on \cg, with data collection occurring approximately six months after its initial release---at a time when adoption in open-source projects was still relatively limited.

In this paper, we replicate and extend our previous study \citep{tufano:msr2024}, by manually inspecting further \newlyInspectedChatGPT more recent---\ie after over two years of the first study---commits, issues, and PRs having traces of self-admitted usage of \cg. This allows to corroborate our previous findings and observe changes in the way in which developers are using \cg for the automation of software-related tasks. On top of this, we also adopt the same study design for GitHub Copilot, thus extending our study to a second AI-based tool, this time specifically designed for developers and integrated in the IDE. In this case, we manually inspected \newlyInspectedCopilot commits, issues and PRs that explicitly referenced the use of Copilot, with the aim of identifying and classifying the software-related tasks developers automated using generative AI. 

\rev{Out of the manually inspected artifacts, we found 343 (\cg) and 281 (\cp) true positive instances, \ie artifacts in which the developers were actually documenting the usage of AI to (partially) automate a software-related task. These have been used to build} two taxonomies of purposes for using \cg and \cp, respectively, in the automation of software-related tasks. ChatGPT's taxonomy features seven root categories and \newCategoriesAllChatGPT categories in total. In general, such a replication confirmed the findings of our MSR'24 study, and the generalizability of the taxonomy we reported there. As per Copilot, we found that its admitted usage is focused on a smaller number of tasks, mostly related to activities requiring strong integration with the IDE and or with other development tools (\eg the issue tracker). This led to a smaller taxonomy, featuring six root categories and \categoriesCopilot categories overall. 

\rev{It must be noticed that our taxonomies represent a biased view of the software-related tasks automated by developers via \cg and \cp. Indeed, we document cases where developers are willing to disclose AI assistance, which could be those in which AI involvement is intentional and socially/technically acceptable to the developer. Thus, our findings do not reflect the complete variety of software-related tasks possibly automated by developers using the subject LLMs.}

\rev{Our findings have implications for both practitioners and researchers. For the former, our study reinforces that LLMs are no longer limited to narrowly scoped coding tasks, but are increasingly embedded across development activities, also acting as brainstorming partners supporting developers in exploring alternative solutions and implementation strategies. Our taxonomies provide an extensive catalogue of software-related tasks that can be supported by LLMs. This deeper integration calls for updated development practices. For example, traditional quality assurance processes may need to be revised being aware of errors which, while rarely performed by humans, may often be made by LLMs (\eg due to artificial hallucination). The increasing importance of quality assurance is also linked to code ownership risks, with developers not always being able to argue for implementation choices that were supported by the AI. }

\rev{For researchers, the tight integration of tools such as Copilot into IDEs makes it progressively harder to distinguish between human-written and AI-generated code, complicating mining studies in need for discriminating the two. This calls for new research methods capturing AI involvement more explicitly. Moreover, the use of LLMs to generate user-facing textual content raises concerns beyond functional correctness, including the risk of producing unwanted, discriminatory, or offensive output, which suggests the need for expanded testing and validation strategies that go beyond traditional software testing techniques. Finally, empirical studies are needed to assess the impact of using LLMs in the automation of the documented tasks, not only when looking at productivity proxies, but also when considering knowledge acquisition, code ownership, \etc}

The remainder of the paper is organized as follows. \secref{sec:design} describes the study definition, and planning. Results are presented in \secref{sec:results}, while their implications are outlined in \secref{sec:implications} and threats to validity are discussed in \secref{sec:threats}. \secref{sec:related} discusses the related literature about generative AI usage in software development, posing particular attention to mining studies. Finally, \secref{sec:conclusion} concludes the paper.

All data used in our study is publicly available \citep{replication}.
\section{Study Design} \label{sec:design}

The \emph{goal} of the study is to unveil the purposes for which LLMs are used to support the development of open-source projects. The \emph{context} consists of \cg and Copilot, as  representative of state-of-the-art LLMs, and of \totalNewlyInspected manually inspected commits, PRs, and issues sampled from open-source projects hosted on GitHub. Our study aims at answering the following research question:

\begin{quote}
\emph{What are the software-related tasks for which developers document the support received by \cg and \cp?}
\end{quote}

We answer this research question by mining, from development artifacts, traces of \cg and Copilot usages. We focus on artifacts for which it is possible to perform keyword-matching queries on GitHub. As such, we search for commit messages, PRs, and issues mentioning \cg or \cp in their textual content. We do not consider GitHub discussions, as we are interested to analyze text directly traceable to software artifacts. Then, we manually inspect \totalNewlyInspected instances with the goal of categorizing the task(s) supported by \cg or \cp, such as code generation or code review. 
On top of that, we classify the instances to determine whether, in that specific circumstance, the LLM-based assistant turned out to be useful or not.
The obtained categories of tasks have then been used to derive two taxonomies of tasks, one for each of the two investigated LLMs. Such taxonomies provide (i) developers with a comprehensive catalog of usage scenarios in which LLMs can be leveraged; and (ii) researchers with software-related tasks which could benefit from automation.

What follows is a step-by-step description of how we structured and conducted our study.

\subsection{Mining Candidate Instances}
\label{sub:mining}
The goal of this step is to identify commits, PRs, and issues in which \cg or Copilot have been likely used to support one or more tasks. False positive instances, \eg instances in which the LLM was mentioned but not actually leveraged, will be discarded in a later stage. 

We started by querying---on January 29, 2025---the GitHub APIs to identify all commits, PRs, and issues containing the word ``\cg'' and/or ``Copilot''. Noteworthy, this query was performed 1.5 years after the analysis conducted in our previous work (early June 2023), which was conducted only six months after \cg initial release.

For commits, the search was performed on the commit message, while for PRs and issues the target was their title and description. This step resulted in a total of 1,179,653 instances, comprising 870,424 commits (622,019 mentioning \cg and 248,405 mentioning \cp), 179,670 PRs (61,131 related to \cg and 118,539 to \cp), and 129,559 issues (93,890 for \cg and 35,669 for \cp). By inspecting the retrieved instances, we noticed a predominance of false positives, mostly due to projects which integrate \cg or Copilot via their APIs. For example, a project may offer features to their users via \cg (\eg a chat bot) rather than using it for automating software-related tasks.  In the case of Copilot, projects involving IDE development frequently discussed challenges related to integrating the language model into the IDE environment.

We then performed a first filtering to automatically discard as many false positives as possible. To this aim, we extracted from the collected instances all forward/backward 2-grams and 3-grams containing the word ``\cg'' or ``Copilot''. For example, let us assume that a PR title features the sentence: \emph{Used \cg to implement tests}. In this case, we extract the backward 2-gram ``\emph{used \cg}'' and the forward $n$-grams ``\emph{\cg to}'' and ``\emph{\cg to implement}''.
We then sorted all extracted $n$-grams in ascending order of frequency, and inspected those appearing in at least 0.1\% of the instances, namely 647 instances for \cg and 412 for \cp. We classified each $n$-gram as likely indicating the LLM support in a task, \eg ``\emph{\cg to generate}'', ``\emph{copilot ai generated}'', or as likely indicating false positives, \eg ``\emph{\cg API integration}'', ``\emph{the copilot icon}''. This resulted in a set of 13 relevant $n$-grams for \cg and 26 for \cp. Both lists are available in our replication package \citep{replication}. 

We excluded from the 1,179,653 collected instances those not containing at least one of such $n$-grams, and all those belonging to GitHub repositories having less than 10 stars in an attempt to filter out toy projects. Finally, we removed duplicates, such as duplicated commits due to forked repositories, obtaining a final set of 35,572 instances, distributed as summarized in \tabref{tab:relevantInstances}.

\begin{table}[ht!]
    \centering
    \caption{Commits, PRs, and issues matching at least one of the $n$-grams likely to indicate the usage of \cg or Copilot in the automation of a software-related task}
    \begin{tabular}{l|rr}
    \toprule
    {\bf } &  {\bf ChatGPT} & {\bf Copilot}\\
    \midrule
    Commits & 627 & 2,637\\
    PRs & 757 & 29,114\\
    Issues & 1,537 & 900\\\midrule
    {\bf Overall} &  {\bf 2,921} & {\bf 32,651}\\
    \bottomrule
    \end{tabular}
    \label{tab:relevantInstances}
\end{table}

\subsection{Manual Analysis and Taxonomy Definition}
\label{sub:collection}
We conducted a manual analysis on a statistically significant sample, selected to ensure a 95\% confidence level with a $\pm$5\% margin of error~\citep{rosner:brooks2011}. For \cg this means obtaining \newlyValidChatGPT valid instances representative of the instances summarized in \tabref{tab:relevantInstances}. To achieve this, the sample was stratified based on the relative frequency of commits, PRs, and issues. Specifically, 53\% of the instances in the sample must be issues, 21\% commits, and 26\% PRs. Note that, to obtain such a sample of valid instances, we had to manually inspect a much higher number of instances (\newlyInspectedChatGPT) since, as explained later, we had to discard false positives, \ie instance in which \cg was mentioned but not used to automate software-related tasks.

We applied the same procedure to the \cp's commits, issues, and PRs. However, upon inspecting some of the instances we quickly realized that several of them could be automatically classified in terms of software-related tasks developers were automating using Copilot. This was due to the usage of pre-defined Copilot templates documenting, for example, when Copilot was used to  generate the summary of the PRs. In this case, a standard text ``\emph{generated by copilot at [commit-id]}'' was present. To give an idea of the magnitude of this automated classification process, we classified through it 21,933 PRs in which Copilot was used to generate the PR description. We will discuss in \secref{sec:resultsCopilot} all the textual patterns which led us to automatically classify instances, since they still represent actual usages of Copilot for the automation of software-related tasks. However, we decided to exclude them from the sample to inspect for the manual analysis. 
 In total, 30,356 instances (out of the 32,651 we matched using $n$-grams) were automatically classified. Thus, we targeted a statistically significant sample of instances (95$\pm$5\%) to manually inspect from the remaining 2,295, \ie a stratified sample of \newlyValidCopilot instances. As we will clarify in \secref{sec:resultsCopilot}, despite inspecting \newlyInspectedCopilot instances, we did not reach our target due to the very high percentage of false positives, thus stopping at \finalCopilot.

The goal of the manual analysis was to characterize within each instance the task(s) (partially) automated using the subject LLMs. Six authors (from now on evaluators) were involved in the manual inspection. Each instance has been independently inspected by two evaluators. The whole process was supported by a web app we developed that implemented the required logic and provided a handy interface to categorize the instance. For each instance, the evaluator was presented with: (i) the metadata as returned by the GitHub APIs; (ii) the $n$-gram that was matched in that specific instance, \eg ``\emph{\cg to generate}''; and (iii) the link to the instance on GitHub for an easier inspection. 

The categorization required the assignment of one or more labels to an instance, describing the automated task(s), such as \emph{refactoring code}, and/or \emph{write documentation}. In case the manual inspection revealed that the LLM was not used to automate software-related tasks, the instance was discarded. Also, the evaluator could add a note to explain whether or not the instance was considered useful by developers, assuming this was possible to be inferred from the discussion.

We used, as starting point the list of tasks elicited in our previous work \citep{tufano:msr2024}. However, evaluators could introduce new labels when needed. When a new label was added, it became available, through the web app, to the other evaluators. While this goes against the notion of open coding, this helps to reduce the chance of multiple evaluators defining similar labels to describe the same task while not introducing a substantial bias in the process. 

Once we reached the set targets for both \cg and Copilot, we solved conflicts. We found conflicts in 43\% of the cases for \cg and 31\% for Copilot. While such percentages may look high, this can be easily explained by two design choices. 
The first is the possibility given to evaluators to add new categories, implying the definition of semantically equivalent but different labels to describe an automated task, such as \emph{create tests} \emph{vs} \emph{test writing}, that would generate a conflict. The second concerns our conservative definition of conflicts: We considered an instance as a conflict if two evaluators assigned a different set of labels to the instance, even if the two sets partially overlapped. Conflicts also arose if one of the two evaluators discarded the instance as a false positive while the other labeled it. Each conflict has been inspected in pairs by two additional evaluators, who discussed and solved it. The \totalNewValid classified instances (\newlyValidChatGPT for \cg and \finalCopilot for \cp) belong to 492 different projects. 

The labels defined through the above-described process have been used to build two hierarchical taxonomies of software-related tasks for which \cg and \cp provided (partial) automation. Two of the authors created a preliminary version of the taxonomy which has then been refined in two rounds by collecting the feedback of all authors involved in the labeling. 

In the following, we report the results by (i) discussing and comparing the taxonomies obtained for \cg and \cp; (ii) discussing differences between the previous \cg taxonomy \citep{tufano:msr2024} and the new one; and (iii) reporting, for the taxonomy high-level categories, the percentage of cases in which the LLM-based assistant was considered useful. \rev{To make the comparison between the two taxonomies (\ie \cg and \cp) simpler, we also provide in \figref{fig:taxonomyMerge} a single taxonomy merging the two.}

\section{Results Discussion} \label{sec:results}

We discuss separately the findings for \cg and \cp in the following subsections. We analyze and discuss them separately, because of their different usage mode. Also, we start from \cg for which we already devised a taxonomy in our previous paper \citep{tufano:msr2024}, and then we complementary discuss the findings for \cp.

\begin{figure*}
	\centering
	\includegraphics[width=1\linewidth]{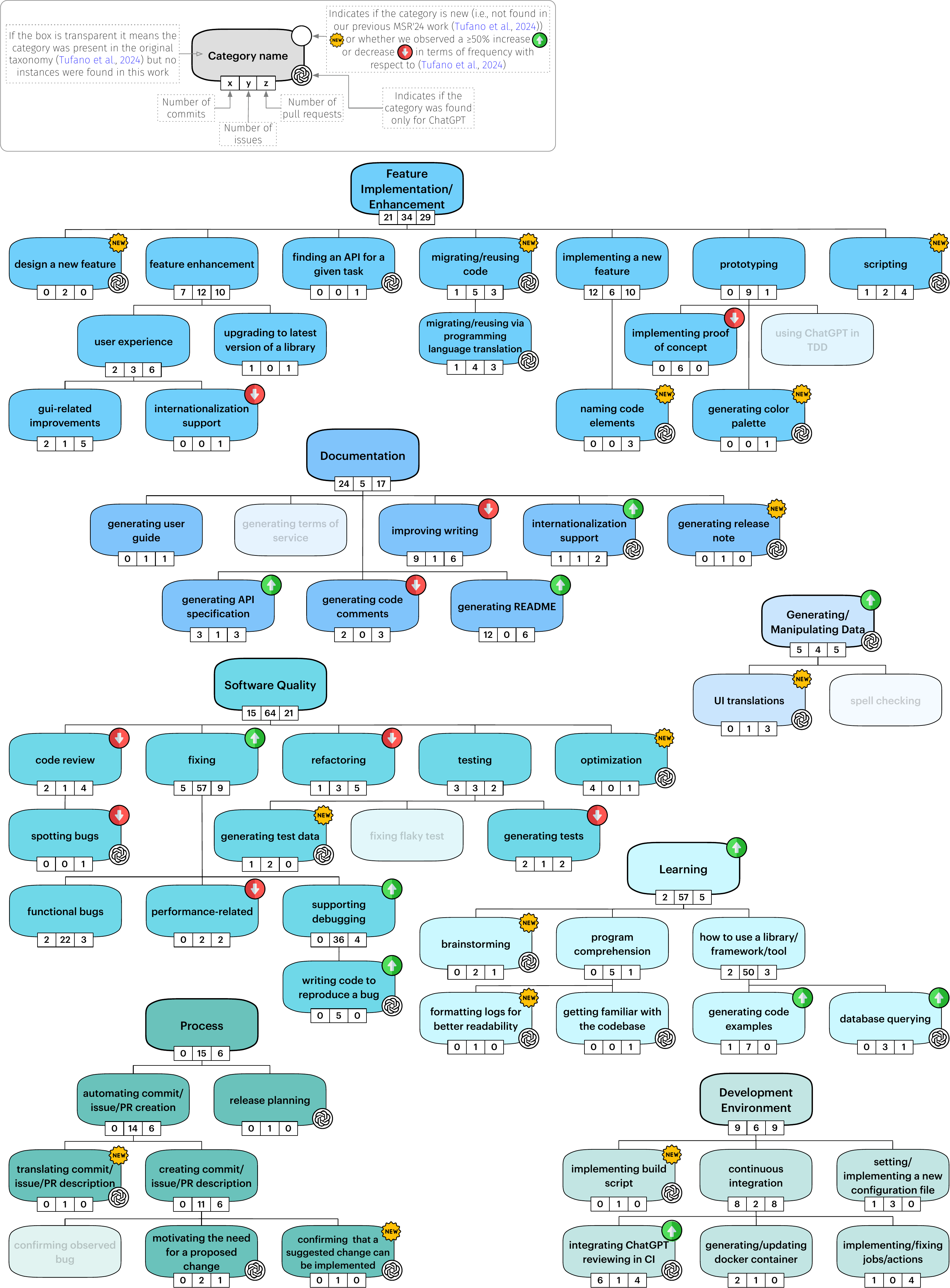}
	\caption{\rev{Taxonomy of tasks for which developers self-admitted the use of \cg}}
	\label{fig:taxonomyChatGPT}
\end{figure*}

\subsection{Software-related tasks automated via ChatGPT}

\figref{fig:taxonomyChatGPT} depicts the taxonomy of tasks automated via \cg. The taxonomy is composed of seven trees, each grouping together related tasks: \emph{feature implementation/enhancement}, \emph{process}, \emph{learning}, \emph{generating/manipulating data}, \emph{development environment}, \emph{software quality}, and \emph{documentation}. 
The numbers attached to each task $T_i$ indicate, from the right to the left, the number of commits, issues, and PRs  from our sample, in which we found evidence of $T_i$'s automation using \cg. For example, we found a total of 84 instances (21 commits, 34 issues, and 29 PRs) in which \cg has been used to automate the implementation or enhancement of a feature. Note that the sum of the number of instances in all tasks is greater than the total number of valid instances we inspected (\newlyValidChatGPT), since one instance may have required the support of \cg for multiple tasks, such as \emph{generating tests} and their related \emph{code comments}. Also, note that the number of instances in a parent category is not always the sum of the instances in its child categories. For example, consider the \emph{software quality $\rightarrow$ fixing $\rightarrow$ supporting debugging} category: Such a task has been automated in 40 instances (36 issues and 4 PRs) and has one child category named \emph{writing code to reproduce a bug}, automated in five issues. The discrepancy arises from the fact that, in 40 instances, \cg was clearly used to assist in the debugging process; however, only in five of those cases was it possible to further specify the task as aiding in bug reproduction.

\figref{fig:taxonomyChatGPT} also depicts differences between the set of tasks automated via \cg that we found in our MSR'24 study versus those we found in this replication. Categories that were not confirmed in the replication are reported in transparency. This happened only for five tasks: \emph{documentation $\rightarrow$ generating terms of service}, \emph{feature implementation/enhancement $\rightarrow$ prototyping $\rightarrow$ using ChatGPT in TDD}, \emph{process $\rightarrow$ automating commit/issue/pr creation $\rightarrow$ creating commit/issue/pr description $\rightarrow$ confirming observed bug}, \emph{software quality $\rightarrow$ testing $\rightarrow$ fixing flaky test}, and \emph{generating/manipulating data $\rightarrow$ spell checking}. It is important to note that each of these five tasks was documented in only a single artifact in our previous study, indicating that they reflect isolated cases of \cg usage for automation rather than widespread practices.

The 14 new categories that emerged from the replication study are marked with a \includegraphics[scale=0.025]{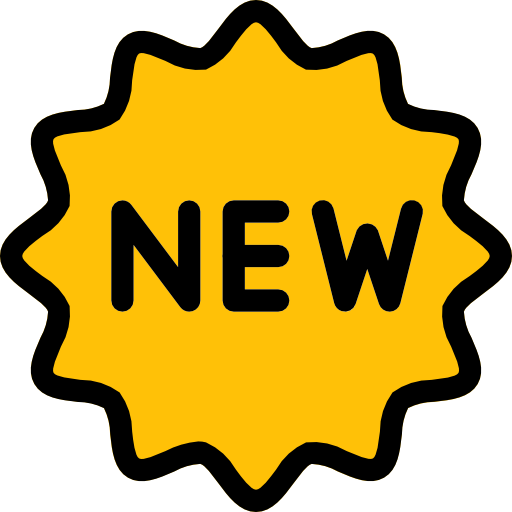} icon in their top-right corner. Also in this case, there are six ``singletons'', namely tasks for which \cg automation was self-admitted only once. 

Categories for which we observed an increase or decrease in their frequency higher than 50\% with respect to our previous study are marked with a \includegraphics[scale=0.025]{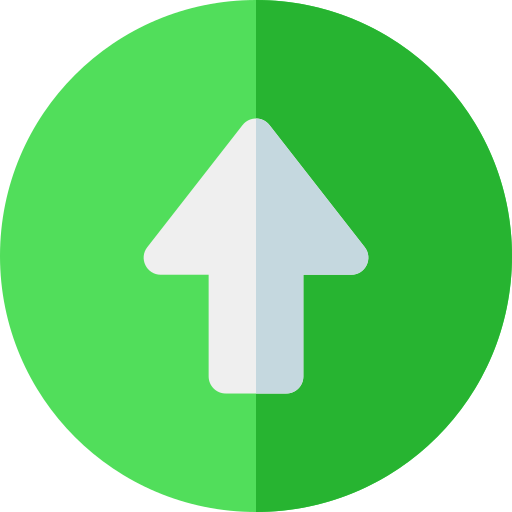}  ($\geq$50\% increase) or a \includegraphics[scale=0.025]{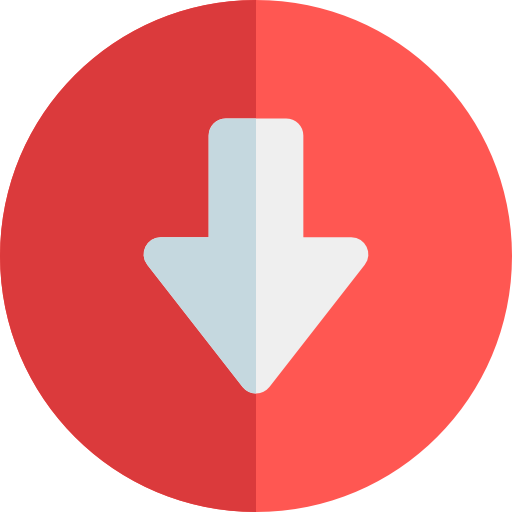}  ($\geq$50\% decrease). These represent tasks for which we observed a substantial change in popularity. However, it is important to highlight that these variations must be evaluated in the context of the number of artifacts related to the task, \ie the three numbers reported at the bottom of each category. What we mean is that a 50\% increase could happen if a task moved from 4 relevant documents to 6. However, this does not have any sort of strong implications when it comes to the observed trends. The differences between our new findings and our original work \citep{tufano:msr2024} are discussed in details in \secref{sub:differences}.

Finally, categories of tasks that we found to be automated via \cg but not via Copilot (its taxonomy is shown in \figref{fig:taxonomyCopilot} and discussed in \secref{sec:resultsCopilot}) are labeled with the \cg logo.

The results' discussion is organized around the seven main categories of automated tasks by reporting qualitative examples and discussing implications for practitioners (see \faLaptop~icon) and researchers (\faLightbulbO). We also showcase \cg's limitations when used for the automation of the related tasks (\faWarning). For the sake of brevity, we do not discuss all \categoriesAll categories in our taxonomy, but only the main ones. However, our replication package \citep{replication} provides the complete dataset reporting, for each category, the instances assigned to it.

\subsubsection{Feature implementation/enhancement}
This category features tasks related to the usage of \cg as a support for implementing and enhancing software features. \rev{We found 84 artifacts (21 commits, 34 issues, and 29 PRs) documenting self-admitted usage of \cg for such a family of tasks.}

We start by commenting two of its related, yet different, sub-categories: \emph{implementing a new feature} and \emph{prototyping}. The \emph{implementing a new feature} sub-category refers to the usage of \cg as a support to implement a specific \emph{part} of a feature that the developer is working on. This means that the developer delegates the implementation of a specific functionality, which is then manually integrated with the rest of the code needed for the feature. An example is a commit from the \texttt{CosmosJourneyer} project\footnote{\url{https://github.com/BarthPaleologue/CosmosJourneyer/commit/26dbb7ca64bc0f10c416c68a9fd255196bc03e67}}, in which the author states: ``\emph{credit \cg for initial version + me for the final version with arbitrary channel merging}''.

The \emph{prototyping} sub-category refers to the usage of \cg as a way to quickly implement either (i) the whole starting version of a project, on which developers can work and build on top, or (ii) a complete feature that can be used as a starting prototype for reasoning about the addition of the new feature in the project. An example of the first scenario is issue \#3529 from the \texttt{micro} project\footnote{\url{https://github.com/zyedidia/micro/issues/3529}}, where the user explains that they had recently started using \texttt{micro} (a text editor) and noticed the lack of Copilot integration. As a result, they developed a complete plugin to enable Copilot support within \texttt{micro} and proposed it to the project maintainers. The interesting part of this issue lies in the following paragraph: ``\emph{This code has been built by 99\% from \cg as I never wrote a line of lua in my entire life}''. As we already observed in our original study \citep{tufano:msr2024}, it is not uncommon to find issues and PRs in which the contributor explicitly states to be unsure about what was accomplished using \cg, or even  declared that they were completely unfamiliar with coding while contributing PRs or issues. Such cases confirm how LLMs facilitates developers' onboarding in software projects \citep{maider24}. 
\faLightbulbO~At the same time, this finding may impact studies involving contributors of open-source projects, \eg studies on newcomers similar to those of  \cite{SteinmacherCTG16} or  \cite{ZhouM10}. That is, studies aimed at profiling open-source project contributors should, in future,  be careful about surveying developers that have only submitted AI-generated contributions, as they may not be representative of the target population. Clearly, the development landscape may also significantly change, as contributors mainly relying on AI when submitting code could become the norm.

\faLaptop~When it comes to practitioners, there are two main aspects to consider. First, similarly to what happens when defining onboarding and contribution guidelines in open-source projects \citep{asfcontrib,eclipsecontrib}, it may be desirable to define guidelines about contributing with AI-generated code, \ie a project may decide to only welcome contributions from \cg by users that are confident in assessing the correctness of the generated code. Second, projects may need to adapt the code review process, \eg relying less on (semi-) automated code quality check when AI-generated contributions come from users having little or no programming expertise. 

Concerning the second usage scenario for prototyping, \ie using \cg to implement a complete feature, a concrete example is issue \#4 from the \texttt{blog} project\footnote{\url{https://github.com/Kerollmops/blog/issues/4}}, where developers are discussing about memory-related optimizations via porting to LMDB (Lightning Memory-Mapped Database). While discussing the attempt done, the user wrote: ``\emph{A fun fact is that it is the first time in a long time that I have written C++, and the first time I asked \cg to code for me because I was not confident in doing C++ and feared falling into some segfault. [...] The chatbot's code was mostly effective, but it omitted a critical empty vector check, which led to a segmentation fault}''. Such an artifact points to \faWarning~ risks related to the ownership and understanding of code contributed via \cg, especially when it is used to contribute complete features. Indeed, this is one of the scenarios in which, without a proper understanding of the generated code, it would have been risky to just use the LLM recommendation. The problem of code ownership for LLM-generated code was well-summarized in a comment of a PR\footnote{\url{https://github.com/typescript-eslint/typescript-eslint/pull/6915}} we inspected in our MSR'24 paper: ``\emph{[$\dots$] I want to make something clear about code suggestions done by \cg: deferring to an AI bot is not the same as code ownership [$\dots$] the idea that an author puts some code into a commit and sends it means they should have an intellectual understanding of it. PR authors should own the code they send -- ownership in the sense of being able to advocate for the code}''.  

Interesting, prototyping via \cg does not only mean code generation, but also collateral aspects such as the choice of the color schema to use in a User Interface (UI)\footnote{\url{https://github.com/ethdebug/format/pull/44}} (category \emph{feature implementation/enhancement $\rightarrow$ prototyping $\rightarrow$ generate color palette}).

The most popular sub-category is \emph{feature enhancement}, in which \cg is used to enhance an already existing feature. Some of these enhancements are generic, and have not been further specialized in \figref{fig:taxonomyChatGPT}. These include, for example, a case in which \cg has been used to improve a regular expression\footnote{\url{https://github.com/DontShaveTheYak/cloud-radar/pull/238}}, with the developer commenting: ``\emph{I used \cg for the regex. It took many tries before it gave me a working one. When I asked it why it took so long for us to derive the correct answer [...] it was able to clearly articulate how we miscommunicated}''. \faLaptop~This PR highlights the increasing importance of prompt engineering skills software developers must acquire to effectively exploit LLMs in their daily activities. 

Other cases are more specific, and have been consequently classified in child categories of \emph{feature enhancement}. 
For example, developers are using \cg for migrating code across programming languages: \emph{migrating/reusing via programming language translation} (8). \faLaptop~Practitioners relied on \cg to automate the translation of complete features across different programming languages, unlocking new opportunities for reuse that were not imaginable in the past. A concrete example is issue \#574 from \texttt{LangChain4j}\footnote{\url{https://github.com/langchain4j/langchain4j/issues/574}}. This project is the Java version of LangChain, a framework for building LLM-powered applications originally developed in Python. Specifically, the contributor wrote: ``\emph{The original (python based) LangChain project has a MarkdownHeaderTextSplitter [...] would be nice to have this as well in LangChain4j [...] I used \cg to translate the Python code to java (including the Unit Test). It might not conform to all the coding standards of the project, but it does the job and tests pass}''. Impressive in this example are two things. First, that the developer managed to translate a total of $\sim$500 lines of code automatically. Second, that the LLM managed to seamlessly translate both production (200 LOCs) and test (300 LOCs) code, allowing a first ``correctness'' check by the developer thanks to the translated tests. \faLightbulbO~Our findings confirm the relevance of research targeting the automated translation of software across programming languages, using statistical methods before \citep{nguyen:ase2014,nguyen:icse2014}, and LLMs more recently, where a challenge is the translation towards low-resource languages \citep{10.1145/3689735}.
\faLaptop~Also, while \cg seems to be able to generalize across several languages, with translations including \texttt{bash} to \texttt{bat}\footnote{\url{https://github.com/apalache-mc/apalache/pull/2980}}, PHP to Python\footnote{\url{https://github.com/Flipnote-Collective/kwz-parser/pull/10}}, Javascript to Python\footnote{\url{https://github.com/Pirate-Weather/pirate-weather-code/issues/52}}, Java to Python\footnote{\url{https://github.com/mikey0000/PyMammotion/issues/4}}, Javascript to PHP\footnote{\url{https://github.com/maxi-blocks/maxi-blocks/issues/4386}}, \etc \faWarning~recent research has pointed out perils of ML-based language translation \citep{malyala2023mlbased}, especially because the translation may not take into account that different programming languages may follow different programming paradigms, such as object oriented \emph{vs} functional, and the result could just be ``Java with a Python syntax'' or something similar. However, we did not find evidence of such a problem in the instances we inspected and it would be interesting to re-assess the extent to which these translation problems still occur when considering modern LLMs. 

Another popular sub-category of \emph{feature implementation/enhancement} concerns improvements to the \emph{user experience}, such as GUI-related enhancements or improvements to the internationalization support. As for the GUI-related improvements, an interesting example is PR \#313 from the \texttt{stampy-ui} project\footnote{\url{https://github.com/StampyAI/stampy-ui/pull/313}}, in which the contributor proposes a fix to better support dynamic heights, \ie ``\emph{advice from \cg 4 helped}'', with another developer commenting: ``\emph{I don't know what this is doing, but it looks sane, the site seems to work, so approving}''. \faWarning~Again, this PR points to the notion of code ownership, which may not always be there for LLM-generated (and merged) code. As per the usage as internationalization support, \faLightbulbO~ researchers working on detecting and fixing internationalization issues should be aware of the fact that \cg could produce mistakes different from those typically committed by developers who manually implement internationalization.

The \emph{feature implementation/enhancement} category includes five new sub-categories of task which we did not document in our previous paper \citep{tufano:msr2024} --- see \figref{fig:taxonomyChatGPT}. The \emph{scripting} category (7 artifacts) is the most popular among those five, and it was kind of expected considering the growth in popularity that \cg experienced among those who are not IT experts in the time period between the two mining studies we performed. We found developers contributing different types of scripts, aimed for example at extending the features offered by an open-source project, such as helper scripts to import benchmarks from another project to the one in which the script has been contributed\footnote{\url{https://github.com/linkwarden/linkwarden/issues/123}}, or at managing the creation of branches/PRs in a project\footnote{\url{https://github.com/zorn/franklin/pull/151}}. 
Another sub-category that emerged from the new sample is \emph{design a new feature}, in which \cg has been used to take decisions about how to implement a given feature rather than as an automated code generator. In issue \#83 from the \texttt{policyengine-uk-data} project\footnote{\url{https://github.com/PolicyEngine/policyengine-uk-data/issues/83}}, the contributor discusses a problem they have with cumulative distribution functions generated by the project which may be invalid. Then, he wrote: ``\emph{After asking \cg for some ideas, I think a promising approach could be first synthesizing a probability density function from the quantiles, smoothing it with a kernel density estimator, then integrating it to a cumulative distribution function}''. \faLaptop~This shows the potential usefulness of tools like \cg even as a brainstorming tool to steer towards the best implementation solution.

We conclude our discussion of the \emph{feature implementation/enhancement} category by commenting on the successful/unsuccessful cases we observed in the artifacts assigned to it. This data must be interpreted knowing the existence of an observational bias, \ie many unsuccessful applications are likely not documented. For this category, we found 80\% of the observed \cg automations to be successful, 11\% to be unsuccessful, and 9\% for which we could not derive any positive or negative outcome. The unsuccessful cases were often due to buggy code produced by \cg, see \eg PR \#1290 from the \texttt{Uplink} project\footnote{\url{https://github.com/Satellite-im/Uplink/pull/1290}}. As said, this is very likely to be an underestimation of problematic code generated by \cg, \faLightbulbO~supporting empirical investigations aimed at assessing the quality of LLM-generated code \citep{yetistiren2022assessing,siddiq2024quality}.

\subsubsection{Process}
The \emph{process} category groups instances mentioning the usage of \cg to support activities related to the development process, \eg \emph{release planning}, or the automation of steps needed to \emph{create commits, PRs, issues}, \eg  generating a PR description. \rev{Overall, we found 21 artifacts (15 issues and 6 PRs) with self-admitted usage of \cg for the automation of process-related tasks.}

As we already observed in our previous study \citep{tufano:msr2024}, there are a few cases in which \cg is used to elicit ideas (and consequently requirements for new features) on how to improve a software project (\emph{release planning} label in our taxonomy)\footnote{\url{https://github.com/GFW-knocker/gfw_resist_tls_proxy/issues/49}}. Both in our original taxonomy and in this replication we found only  a single data point showing this application of \cg. However, this shows how LLMs can go substantially further than what state-of-the-art tools supporting release planning are able to do. The latter usually mine data, such as app reviews \citep{CiurumeleaSPG17,PanichellaSGVCG15,ScalabrinoBRPO19}, to help developers summarizing the customers' feedback and come up with aspects to improve in the software. \faLaptop~\cg does not require any sort of data mining on the developer's side and, as visible from the issue we inspected\footnote{\url{https://github.com/greenshot/greenshot/issues/499}}, can be queried for general ideas about what to improve in a software or even on how to improve a specific feature/quality aspect of the software. These span from improvements which will mostly have an impact on software quality 
to recommendations pointing to possible new features to be implemented.
\faWarning~One clear limitation is that requirement crowdsourcing may require up-to-date sources of information, \eg recent information about the features that competitive software implements, which \cg may not have. However, the most recent models at the core of \cg can handle extremely large inputs and support web search as a feature. \faLaptop~This may allow practitioners to either input information about competitive projects, or just pointing \cg to the list of main competitors via \eg links to their websites or repositories.

In 17 instances developers used \cg to automatically generate a PR/issue description, \eg issue \#759 from the \texttt{extension} project\footnote{\url{https://github.com/YouTube-Enhancer/extension/issues/759}} --- \emph{process $\rightarrow$ automating issue/pr creation $\rightarrow$ creating issue/pr description}. This sub-category warrants clarifications due to a significant change from our previous study. In our previous taxonomy \citep{tufano:msr2024}, it was named \emph{creating commit/issue/pr description}, as we found multiple instances where \cg was used to automatically generate commit messages. While this does not necessarily mean that \cg is no longer used for generating commit messages, as we will discuss in \secref{sec:resultsCopilot}, this task now appears to be primarily handled by Copilot. \faLightbulbO~This shift suggests that developers may favor IDE-integrated LLMs for tasks that (i) 
are performed frequently --- commits occur far more often than issues or PRs ---; and (ii) benefit from readily accessible contextual information within the IDE, such as the associated code changes, thereby minimizing context switching and the need to write prompts manually. 

Our taxonomy also features two interesting specializations of the \emph{creating PR/issue description} sub-category. The first concerns a scenario in which \cg has been used to confirm that a suggested change can be implemented: ``\emph{[proposed feature description] I asked \cg and this is totally possible to do without too much effort}''\footnote{\url{https://github.com/MobiFlight/MobiFlight-Connector/issues/1533}}. The feature has then been implemented in a subsequent PR. The second represents instances in which \cg was used to better motivate a proposed change. An example is PR \#151 from the \texttt{psn-api} project\footnote{\url{https://github.com/achievements-app/psn-api/pull/151}}, being a code refactoring adding type declaration to package exports. The PR description includes text from \cg explaining why such a refactoring is important: ``\emph{Adding types to package.json exports helps typescript find the type declarations of your package [...] Since your type declarations were there, TypeScript was able to find them, and you didn’t see any error [...] However, it’s generally better to specify all your module's entry points [...] this way, you're not relying on TypeScript's fallback behavior, which might not work in all environments}''.

All 21 artifacts in which we found applications of \cg for ``process automation'' were successful, in the sense that developers were happy with the recommendations provided by the LLM. This was not the case in our original study, in which we found a few cases of AI hallucinations\footnote{\url{https://github.com/pizzaboxer/bloxstrap/issues/224}}.


\subsubsection{Learning}
The \emph{learning} category primarily includes issues \rev{(57 out of 64 artifacts)} that stem from difficulties users encounter while working with a specific library, framework, or tool—accounting for 50 out of the 64 artifacts. In these cases, users often describe their efforts to resolve the issue by consulting \cg. For example, one user noted: ``\emph{I googled and asked \cg about advices, but I cannot see what could have gone wrong during the integration.}''\footnote{\url{https://github.com/home-assistant/core/issues/135369}} \faWarning~Confirming our former findings \citep{tufano:msr2024}, this is a category of tasks for which \cg showed clear limitations, primarily due to AI hallucinations. In 73\% of artifacts we inspected related to learning how to use a library,framework, or tool, the support provided by \cg was incorrect, leading users to report negative feedback when opening the issue, \eg 
``\emph{I tried to use \cg to get some help with avc codec parsing [...] but I'm not luck for now}''\footnote{\url{https://github.com/Vanilagy/mp4-muxer/issues/11}}. \faWarning~For learning and solving issues related to the usage of libraries, frameworks, and tools, the more classic Q\&A websites such as Stack Overflow seem to still be a better solution as of today, as confirmed by \cite{10.1145/3613904.3642596}.

The \emph{learning} category also features eight instances in which \cg has been used for code comprehension. In these cases, the LLM provided useful support to developers, by reformatting log messages to improve their readability\footnote{\url{https://github.com/freelawproject/courtlistener/issues/4646}} or clarifying whether specific changes would impact the code behavior\footnote{\url{https://github.com/opentypejs/opentype.js/pull/572}}.

Finally, a new sub-category that emerged is \emph{brainstorming}, referring to cases in which \cg is used as a mean to collect perspectives on something which must be better understood or that is up for discussion. For example, in issue \#1292\footnote{\url{https://github.com/huggingface/blog/issues/1292}} the developer is reporting a possible contradiction in Hugging Face documentation and writes: ``\emph{I tried to research the RL fine-tuning process a bit and asked \cg to brainstorm with me about this contradiction. From what I've gathered [...]}''. \faLightbulbO~Such an usage of \cg, especially when viewed alongside other brainstorming-related tasks discussed earlier, \eg release planning, highlights the automation opportunities that LLMs offer to software developers---even for tasks that require creativity and complex reasoning.

\subsubsection{Generating/manipulating data}
Developers use \cg to easily \emph{generate/manipulate data}, \rev{even though such a usage scenario is not so frequent, with an overall of 14 artifacts providing evidence of it (5 commits, 4 issues, 5 PRs).}

The variety of data involved in this category includes LLMs' prompts (``\emph{changed prompts with help of \cg}''\footnote{\url{https://github.com/clp-research/clembench/commit/4d3c6c83ceb}}), documentation-related translations (``\emph{Here is the translation of the LibreOffice extension, I used \cg for it, so I can only assure that Portuguese Brazilian and Portuguese European are correct}''\footnote{\url{https://github.com/writingtool-org/writingtool/issues/10}}), and strings appearing in the UI for user interaction\footnote{\url{https://github.com/flytegg/ls-discord-bot/pull/243}}. \faWarning~It is worth highlighting here possible risks in using LLMs for generating strings used for interactions with the user, since they may introduce biases \citep{ChakrabortyM0M20,ChakrabortyMM21} or produce discriminatory or offensive content. Moreover, LLMs could be vulnerable to adversarial attacks, potentially leading to unintended outputs, as it has been shown for other recommender systems \citep{NguyenSRPR21}. Nonetheless, all instances in the \emph{generate/manipulate data} category were classified as successful applications of \cg.

\subsubsection{Development environment}
This category groups instances where \cg has been used to support and (partially) automate activities related to the \emph{development environment}. \rev{A total of 24 artifacts (9 commits, 6 issues, and 9 PRs) provide evidence of such a \cg usage.} The most popular application of \cg in this context is its \emph{integration as reviewer in the continuous integration and delivery (CI/CD) pipeline}. In this scenario, the idea is to use \cg to comment about contributed code and identify bugs and/or suboptimal implementation choices\footnote{\url{https://github.com/taikoxyz/taiko-mono/pull/13786}}.
\faLaptop~The \cg-based review is usually integrated in the CI/CD pipeline and aims at providing a first quick feedback to the contributor, supporting human reviewers. Furthermore, \cg is usually combined with classic linters looking for quality concerns and assessing test coverage. \faLightbulbO~This application reinforces the significance of recent research efforts focused on automating code review tasks \citep{tufano:icse2022}, positioning \cg as a potential baseline for future comparisons. Future research should also consider how to properly leverage LLMs to obtain code reviews in line with an organization/project's own coding styles and guidelines. \faWarning~A clear issue is the need for passing code to \cg, which may not be acceptable (and even forbidden) in industrial environments. In such cases, approaches leveraging local LLMs, possibly combined with specific fine-tuning or Retrieval Augmented Generation architectures are to be preferred. 
 
Other tasks automated via \cg concern the \emph{implementation/fixing of jobs/actions} in CI scripts and the \emph{generation/updating of docker containers}. As for the former, developers managed to implement entirely GitHub actions, \ie workflows that automate software development processes within GitHub repositories, by asking \cg: \emph{This PR adds a GitHub Action that will add the ``No changelog entry needed'' label to PRs that include ``Update pinned requirements'' or ``autoupdate'', or contain the pattern ``Bump * from * to *''; I used \cg to generate the GitHub Action}\footnote{\url{https://github.com/PlasmaPy/PlasmaPy/pull/2249}}. Also, \cg supported the refactoring of deployment scripts, cleaning them from unneeded code\footnote{\url{https://github.com/loculus-project/loculus/pull/655}}. As for the latter, we found both successful and unsuccessful applications of \cg. For example, in commit \#070a74f of the \texttt{larentals} project\footnote{\url{https://github.com/perfectly-preserved-pie/larentals/commit/070a74f3b158eaf570f9ea9b8f6243331f2dc9c1}}, the developer is fixing a syntax error in a Docker file, commenting: ``\emph{This is why you don't use \cg for Docker command syntax}''. While \cg can be helpful for drafting CI/CD scripts, LLMs outcome suffers from their inability to be continuously retrained to be updated. Given that CI/CD technologies are relatively new and evolve rapidly, \cg may sometimes provide outdated or ineffective recommendations. \faWarning~LLMs might not be suitable in rapidly evolving contexts such as young technologies, programming languages, \etc Indeed, we marked as successful $\sim$60\% of the instances in the \emph{development environment} category. While this may look as a positive finding, it is worth remembering the likely observational bias characterizing this analysis, \ie higher chance of observing successful applications of \cg rather than documented failures. \faLightbulbO~In rapidly evolving contexts, it is possible that smaller, specialized models that can be quickly retrained might be more suitable and reliable.

\subsubsection{Software quality}
Confirming what observed in our MSR'24 paper, \emph{Software quality} is the largest category in our taxonomy in terms of number of instances \rev{(100 artifacts, out of which 15 commits, 64 issues, and 21 PRs)}. This confirms that developers largely leverage \cg for automating tasks related to software quality improvement. However, we found major differences in the popularity of the sub-tasks for this category. Indeed, we observed a decrease in usage of \cg when it comes to recommending refactoring operations ($-70\%$) accompanied by an increase in its usage to support bug-fixing activities, with the \emph{fixing} sub-category that went up $+76\%$ in popularity. As already observed for the automated generation of commit messages, such a shift over time may be due to better in-IDE tools developed in the meanwhile, which may be better suited for tasks such as code refactoring.

We found 41 artifacts in which the usage of \cg to \emph{support debugging} was documented and, in some of them, developers also used the LLM to fix the spotted bug. A concrete example is issue \#639 from the \texttt{exo} project\footnote{\url{https://github.com/exo-explore/exo/issues/639}}, in which the developer wrote: ``\emph{I was getting errors like Error processing prompt: Invalid control character at: line 563 column 62 (char 47769) while running exo [...] With the help of ChatGPT (o1), I tracked it down to a bug in hf\_helpers.py}''. The LLM went even further by explaining the bug (\emph{As \cg says, after that code block, there's no return or raise if downloaded\_size != total\_size; the function simply falls through to this part}) and recommending a fix for it (\emph{Here is the fix suggested by ChatGPT o1: [23-line diff]}). While this was a successful application of \cg as a support to debugging and bug-fixing, we also found counterexamples in which the LLM was providing wrong recommendations. As an example, in issue \#151 of the \texttt{zotify} project\footnote{\url{https://github.com/zotify-dev/zotify/issues/151}}, a contributor reported two errors experienced with the software and then mentioned: ``\emph{Used ChatGPT to troubleshoot and fix both so someone may want to review}''. The answer of one of the project's maintainer was: ``\emph{I recommend to not follow \cg advice. Doing the first causes some songs to keep their temporary name .\_1.mp3 which then causes the other issue. I haven't looked further into the second error but have my doubts that it properly fixes the issue and renames the file}''. There are two important lessons that can be learned from this example. First, \faLaptop~as previously discussed, some open-source contributors may lack the needed expertise to filter out wrong \cg recommendations, requiring extra care in the inspection of the contributed code. Second, \faWarning~the wrong fixes generated by \cg seems to also be the result of a quite limited coding context it was provided with. Indeed, the project' maintainer is pointing to potential drawbacks caused by the fix in other functions, rather than the one subject of the fix. It is possible that with additional context the LLM would have been able to provide a proper fix. However, this is not something we can speculate without knowing the prompt used in this instance. In general, $\sim$50\% of the instances assigned to the \emph{software quality $\rightarrow$ fixing} sub-category have been labeled as ``successful'' applications of \cg, indicating great potential, as well as room for improvement.

Besides the classic \emph{debugging} process, \faLaptop~\cg has also been used to \emph{write code to reproduce a bug} (see \eg ``\emph{I used \cg to create a easy example of this issue}''\footnote{\url{https://github.com/holoviz/panel/issues/6869}}). \faLightbulbO~As of today, there are state-of-the-art approaches supporting the automated bug reproduction \citep{BernalCardenasCHMCPM23,FazziniMBWOP23,ZhaoSLZWKHY22}. This is mainly possible because such approaches are (i) tailored for specific categories of applications, \eg mobile apps, and (ii) can access the whole application code base. LLM-based approaches are now capable, in most cases, of processing entire software repositories (or substantial portions of them) and should therefore be explored and considered as baselines for automated bug reproduction.

In the context of \emph{refactoring} operations, developers used \cg as support to a variety of code transformations, including renaming\footnote{\url{https://github.com/fzyzcjy/flutter_rust_bridge/issues/1381}}, typing-related refactorings\footnote{\url{https://github.com/alkem-io/server/pull/3350}}, and simplifications of the code logic\footnote{\url{https://github.com/ReliaQualAssociates/ramstk/pull/1404}}. \faLightbulbO~While we found a wide variety of refactoring actions automated via \cg, we observed a lack of code transformations involving multiple files, such as extract class refactoring. This is likely due to the lack of \cg integration in the IDE. 
In terms of success, in most cases (77\%) developers were happy with \cg recommendations. A clear example of an unsuccessful application is a PR applying typing-related refactoring, where one of the project maintainers criticized the submitted code contribution:

\begin{quote}

\emph{Unfortunately, you fell into the classic new GPT-developer trap underestimating complexity, effort and risk and taking GPT's response for safe and correct. There are several issues with this PR:}

\begin{itemize}
\item \emph{typings - \{\} is not the same as []. I don't see any signature changes. Also, a simple search of the obsolete type FindOptionsRelationByString still shows 27 hits and this needs to be replaced with FindOptionsRelations};
\item \emph{After replacing these types there are more errors that surface};
\item \emph{refactoring - after changing FindOptionsRelationByString with FindOptionsRelations all places loading relations will need to be refactored, which are quite likely in the hundreds, as merging an object of type X with Y is not the same as merging an object X with Z}
\end{itemize}

\emph{Fixing all this can easily take more than a day and I am not sure this is a day we want to spare this sprint.}\footnote{\url{https://github.com/alkem-io/server/pull/3350}}

\end{quote}

\faWarning~This example highlights again the risk of accepting confident suggestions from \cg, even for code changes which can have substantial impact on the entire code base, with the LLMs ignoring such drawbacks. 
  
Another relevant sub-category is related to the automation of \emph{testing} activities. \cg has been mostly used to generate tests and test data. Tests here may refer to classic test suites, \eg ``\emph{testcode for Bencode function generated by ChatGPT}''\footnote{\url{https://github.com/archeryue/go-torrent/commit/53e37826497}}, as well as, what developers call ``qualitative tests'' aimed at testing a chatbot, \eg ``\emph{Used \cg to help come up with the qualitative tests, which means we are now using LLMs to generate test cases for LLM powered app with a LLM-based testing library}''\footnote{\url{https://github.com/mongodb/chatbot/pull/113}}. As per test data, it depends on the specific tested application, and may be far from trivial. For example, while testing the \texttt{mediawiki-tools-codesniffer}, developers instructed \cg to ``\emph{generate a snippet of highly inconsistent PHP code}''\footnote{\url{https://github.com/wikimedia/mediawiki-tools-codesniffer/commit/c2f2b1282de0f2e9b453e882d5312cd856b24c97}}, to check whether their linter was able to catch the quality issues. \faLightbulbO~Such an application could help even researchers to experiment with techniques supporting bug-fixing, \eg by generating difficult to catch bugs, code smell detection, and quality assurance in general.

\subsubsection{Documentation}
The last popular application of \cg we discuss (46 instances in our sample --- \rev{24 commits, 5 issues, and 17 PRs}) concerns the automated generation of software \emph{documentation}. \cg is used both to write documentation from scratch  as well as to improve existing documentation. In some cases the documentation is generated as a consequence of difficulties experienced by developers in understanding code. For example, in PR \#43 of the \texttt{mem0} project\footnote{\url{https://github.com/mem0ai/mem0/pull/43}}, the contributor wrote: ``\emph{regex is hard to read, so I added some \cg generated documentation to know what's going on}''. Looking at some of the comments contributed by \cg, it is clear how they could improve program comprehension, especially when it comes to developers not experienced with regular expressions:

\begin{quote}

\emph{
This regex identifies consecutive non-alphanumeric characters (i.e., not a word character [a-zA-Z0-9\_] and not a whitespace) in the string and replaces each group of such characters with a single occurrence of that character. For example, ``!!! hello !!!'' would become ``! hello !''.}

\end{quote}

To some extent, \faLaptop~this is further hint into the ability of \cg to support program comprehension. 

Besides code comments, \cg has also been used to generate \texttt{README} files\footnote{\url{https://github.com/cargo-public-api/cargo-public-api/pull/654}}, user guides\footnote{\url{https://github.com/greenshot/greenshot/issues/499}}, API specifications\footnote{\url{https://github.com/hezarai/hezar/pull/124}}, and release notes\footnote{\url{https://github.com/commons-app/apps-android-commons/issues/6025}}. Two interesting observations can be made from these applications of the LLM. First, \faLaptop~in some cases the documentation to write may be subject to specific requirements not easy to fulfil, with an assistant like \cg possibly being very useful. For example, in PR \#654 of the \texttt{cargo-public-api} project\footnote{\url{https://github.com/cargo-public-api/cargo-public-api/pull/654}}, the contributor wrote: ``\emph{Add trademark notice to READMEs to fulfil requirements outlined in the Rust Project's trademark policy, more specifically [link] (I got help from ChatGPT to formulate the notice)}''. Second, \faWarning~while we mostly found successful applications of \cg for generating documentation ($\sim$90\%), also when asking for natural language text it must be considered the possibility that \cg hallucinates, providing what look as confident but wrong recommendations. For example, a contributor proposed the addition of \texttt{pip} commands to the documentation, saying that they \emph{didn't know how to do this so asked ChatGPT}\footnote{\url{https://github.com/tatsu-lab/stanford_alpaca/pull/160}}. However, the contribution was not merged since the provided commands did not actually work.

\subsubsection{Our replication \emph{vs} findings of the previous study}
\label{sub:differences}
Comparing our current findings with those from our previous work, the first notable conclusion is the strong generalizability of our original taxonomy of tasks automated using \cg. Indeed, only five of its 52 original sub-categories have not been confirmed when inspecting the new sample and, as anticipated, all of them were singletons in the original study. 
Also, while we found 14 new categories, again most of them do not represent software-related tasks frequently automated via \cg. Indeed, six of them are singletons, with the most popular one being \emph{feature implementation/enhancement $\rightarrow$ scripting} with seven instances. 

When looking at changes in the popularity of the other sub-categories, it is interesting to discuss cases which (i) experienced substantial changes, and (ii) for which we had a substantial number of instances in the previous taxonomy \citep{tufano:msr2024} and/or in our new taxonomy. In particular, we only discuss tasks for which in at least one of the two taxonomies we had 30 instances documenting their automation (thus allowing to observe meaningful trends).
Given these requirements, the tasks for which we witnessed an increased frequency in the newly analyzed sample are those related to \emph{software quality $\rightarrow$ fixing $\rightarrow$ supporting debugging} ($\times$3.5), and (ii) \emph{learning} (+53\%). Instead, those for which we observed a lower number of artifacts documenting their application are the ones related to \emph{refactoring}, \emph{code review}, and the writing/improvement of \emph{code comments}.
A possible explanation of these increases and decreases is that for tasks such as debugging and technology learning, there is little in-IDE support even for IDE-integrated tools such as Copilot. Indeed, already before the LLM era, help for debugging or for learning how to use libraries, frameworks and tools was usually sought on Q\&A platforms. While LLMs-based assistants are slowly changing this landscape, they still push developers outside the IDE, asking general-purpose LLMs such as \cg. 
Instead, tasks such as refactoring, code review and code comment improvements clearly 
benefit from the additional context present in the IDE, and are well-supported by LLM-based in-IDE tools, as we will discuss in the next section for Copilot. In short, on the one hand the observed results seem to reward in-IDE tools for tasks requiring a heavy and complex context to work properly. On the other hand, general-purpose LLMs---often accessible through a Web-based interface---are still preferred for tasks requiring major reasoning and natural language processing capabilities. 

\subsection{Software-related tasks automated via Copilot}
\label{sec:resultsCopilot}

\tabref{tab:copilot_filter} reports the number of artifacts automatically classified through pattern matching. As discussed in \secref{sub:collection}, many instances admitting the Copilot usage could be automatically identified due to the presence of predefined templates that document its usage. This allowed the automatic classification of 1,212 commits and 28,004 PRs under the \emph{process $\rightarrow$ automating commit/issue/pr creation} sub-category, and 633 commits and 507 PRs under the \emph{software quality $\rightarrow$ fixing}. The latter have been identified by matching the textual pattern ``copilot autofix'', which is a trace left by Copilot Autofix\footnote{\url{https://docs.github.com/en/code-security/code-scanning/managing-code-scanning-alerts/responsible-use-autofix-code-scanning}}, a GitHub code scanning extension.  Copilot Autofix leverages the LLM to suggest fixes for potential vulnerabilities and coding issues flagged during code scanning. 

\begin{table}[ht!]
    \centering
    \caption{Self-Admitted usages of \cp automatically labeled}
    \begin{tabular}{l|rr}
    \toprule
    {\bf Task} &  {\bf Commits} & {\bf PRs}\\
    \midrule
    \emph{process $\rightarrow$ automating commit/issue/pr creation} & \multirow{ 2}{*}{1,212} & \multirow{ 2}{*}{28,004}\\
    \emph{$\rightarrow$ creating commit/issue/pr description} & & \\
     & & \\
    \emph{software quality $\rightarrow$ fixing} & 633 & 507\\
    \bottomrule
    \end{tabular}
    \label{tab:copilot_filter}
\end{table}

\begin{figure*}
	\centering
	\includegraphics[width=1\linewidth]{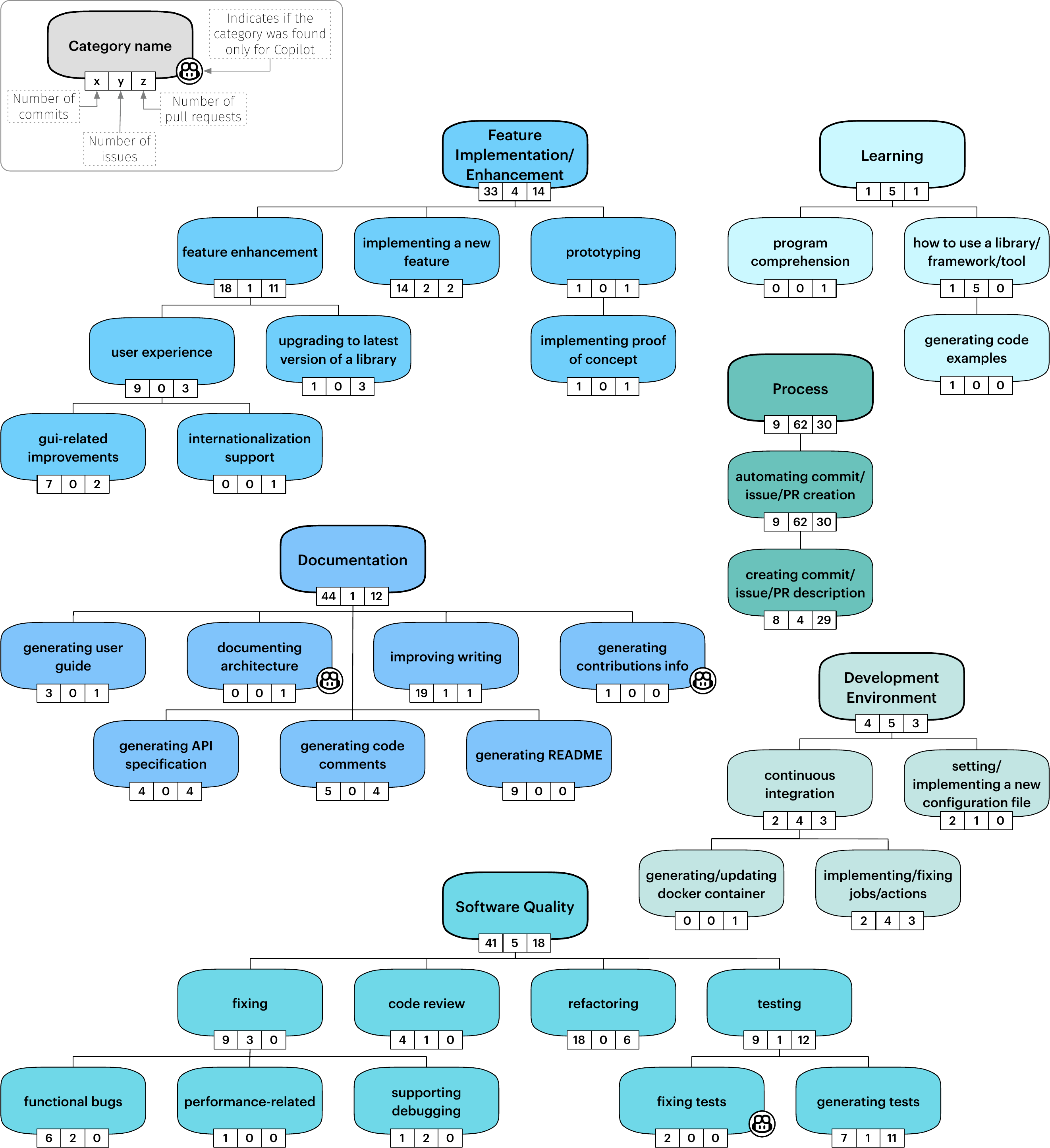}
	\caption{\rev{Taxonomy of tasks for which developers self-admitted the use of \cp}}
	\label{fig:taxonomyCopilot}
\end{figure*}

As for the remaining artifacts (2,295), we targeted the manual validation of a statistically significant sample (\newlyValidCopilot), stratified on the popularity of each type of artifacts in the population, \ie 123 commits, 76 PRs, and 130 issues. However, after manually inspecting \newlyInspectedCopilot of them, we found that the false positive rate (81\%) among the ``issues'' was extremely high, which led us to conclude that further manual labeling would not be worthwhile. Specifically, after inspecting 427 issues only 82 were valid, of which 62 referred to cases in which Copilot was used for automatically creating the issue description. Therefore, we decided to stop the manual validation before reaching the target number of valid issues, as it was unlikely to yield any meaningful insights. 

\figref{fig:taxonomyCopilot} depicts the taxonomy of tasks automated via \cp.  In total, there are \categoriesCopilot categories, 22 fewer than those identified for \cg. This result is not surprising, given that \cg is a general-purpose LLM, whereas Copilot is primarily designed for integration within IDEs, yet it can either be used to simply auto-complete code, or through its chat, which interaction is somewhat similar to \cg (although \cp chat eases embedding code from the IDE in the prompt).

 While six out of the seven root categories observed for \cg have been confirmed (all but \emph{Generating/Manipulating Data}), the \emph{Learning} category of Copilot only features seven instances (2\%), against the 64 of \cg (19\%). This outcome is, once more, expected considering the distinct purposes of the two LLMs, with \cg slowly replacing what a few years ago was Stack Overflow when it comes to learning how to use libraries, frameworks, or tools. While also the \emph{Development Environment} subtree features a limited number of instances (12), this finding is aligned to what observed for \cg, thus not showing major differences.
The remaining four root categories show similar prevalence between \cg and Copilot. However, it must be remembered that we automatically labeled thousands of instances in which \cp has been used to (i) \emph{creating commit/issue/pr description}, and (ii) \emph{fixing} quality issues (see \tabref{tab:copilot_filter}). \faLaptop~Copilot’s integration with GitHub gives it a significant advantage over \cg for tasks that involve interacting with software repositories.

In the following, we discuss these four categories, emphasizing differences in their leaves compared to \cg. Also, we do not evaluate whether \cp's usage was successful, as it was in 96\% of the labeled instances. Instead, we only discuss interesting (but isolated) cases of unsuccessful use.

\subsubsection{Feature Implementation/ Enhancement}
Not surprisingly, \cp is extensively used as a support for implementing new features or enhancing existing ones\rev{, with 51 artifacts reporting self-admitted usages of \cp for this family of tasks (33 commits, 4 issues, and 14 PRs)}. Most likely, this may be due to its IDE integration, and therefore the fact that the generated code is already well-adapted to the current context, \eg in terms of variable names.
However, differently from \cg, it must be noted that it is much more difficult to assess the extent to which Copilot contributed to the code implementation. While in the \cg's instances developers were often explaining the process that led them to the code implementation and the role that \cg played, for Copilot very often we just found short textual patterns (\eg ``Co-authored by Copilot'') in the commit/issue/PR description. This may indicate the simple activation of Copilot as a code completion tool in the IDE, as well as its usage as a chat-based assistant, to which a prompt has been provided. Thus, \faLightbulbO~for researchers mining code from software repositories, it may be difficult to clearly discriminate the automatically generated from the manually written code. At most, we can talk about code written with AI support.

There are a few cases in which a more detailed interaction was documented, such as PR \#21 from the \texttt{localise.travel} repository\footnote{\url{https://github.com/championswimmer/localise.travel/pull/21}}. In this case, the contributor is implementing a new feature, also providing the prompt used for Copilot. \faWarning~Also, we found an interesting PR in which the developer was fixing ``\emph{Copilot generated bugs}''\footnote{\url{https://github.com/dynamiqs/dynamiqs/pull/403}}, stressing once more the need for carefully assessing the LLMs' recommendations. \faWarning~Some developers reported the tendency of Copilot to inject code not really required for the task at hand\footnote{\url{https://github.com/ourjapanlife/findadoc-web/issues/331}}: ``\emph{@anonymized\_dev loves Github Copilot and it has a bad habit of injecting raw strings of code where they shouldn't be. For example, import at the top of .vue files. All random raw lines of code at the top and bottom of files are removed}''. \faLightbulbO~This leads to relevant questions about the impact that LLM-based assistants have on the way developers develop and maintain code, including potential drawbacks in terms of quality of the produced code, as well as of cognitive effort required to quickly assess the received recommendations.

Last but not least, differently from \cg, we did not find usages of Copilot aimed at migrating code across languages, and only two instances in which it was used to prototype a proof of concept. 

\subsubsection{Documentation}
Copilot has been used by developers as a support to produce software documentation, especially when it comes to improvements of already existing documents. Documentation here may refer to code comments\footnote{\url{https://github.com/Mapsui/Mapsui/commit/bb9033dee34}}, user guides\footnote{\url{https://github.com/owid/owid-grapher/pull/4122}}, contribution guidelines\footnote{\url{https://github.com/trillek-team/tec/commit/87a9cd1fd463}}, or architectural documents\footnote{\url{https://github.com/kamranahmedse/developer-roadmap/pull/7082}}. \rev{In total, we found 44 commits, 1 issue, and 12 PRs in which \cp has been used to produce documentation.}

As with code implementation, it can be challenging to distinguish which portions of documentation writing were manually authored, automatically generated by Copilot, or simply assisted by Copilot. \faLightbulbO~Datasets featuring this kind of information, for example by monitoring the developers' activities in the IDE, may allow investigating novel research questions comparing content written with various levels of automation, \eg fully manual, fully automated, AI-supported, across different dimensions.

\subsubsection{Process}
\rev{Process-related tasks supported by \cp deal with the (semi-)automatic creation of commits (9), issues (62), or PRs (30).} The automated generation of PRs description is the task most frequently automated via Copilot, likely due to its integration with GitHub. This is one of those features which have become way more convenient with Copilot rather than with \cg. By looking at the generated summaries, it is clear that Copilot almost solved a problem tackled in several research papers \citep{Liu2019,FANG2022111160}, also for PRs featuring dozens of commits, see \eg PR \#10 from the \texttt{kadeck} project\footnote{\url{https://github.com/basementdevs/kadeck/pull/10}}, featuring 30 commits, with the developer writing: ``\emph{this PR has so many changes that I left it to the Copilot to generate the description}''. 
In another case we found an issue description generated ``\emph{using Copilot from Messages in the Buefy Discord}''\footnote{\url{https://github.com/ntohq/buefy-next/issues/297}}, \faLightbulbO~supporting the recent line of research looking at instant messaging platforms as a source for software documentation \citep{raglianti:icpc2022}. In this specific case, the mining of the Discord messages reveals the rationale motivating the opening of the issue.

\subsubsection{Software Quality}
\rev{This category groups all tasks dealing with the automation of activities aimed at improving code quality. We found 64 artifacts in which developers self-admitted the usage of \cp for the automation of the related task, most of which are commits (41).} Overall, Copilot seems to be preferred to \cg when it comes to the improvement of software quality. Again, this is likely due to its integration in the IDE and the ``knowledge'' of the code base. Indeed, besides the 64 instances from our taxonomy documenting its usage for quality improvement, we must also remember the 633 commits and 507 PRs in which it was used for fixing code quality issues. 

Comparing the Copilot and \cg taxonomies, it is clear that activities such as \emph{refactoring} and \emph{testing} are more frequently supported using Copilot than \cg. In a commit of the \texttt{operaton} project\footnote{\url{https://github.com/operaton/operaton/commit/52653be0f1a}}, the developer used Copilot to refactor the test code (``\emph{refactor to use assertThatThrownBy() for checking code for exceptions thrown}''), also documenting the used prompt.
\faLaptop~\faLightbulbO~This example confirms the importance of prompt engineering also for IDE-integrated LLMs such as Copilot. 

In other cases, the production code was subject of the refactoring, with Copilot optimizing code\footnote{\url{https://github.com/gitextensions/gitextensions/pull/11268}}: ``\emph{This is partly a test of how you can use Copilot to optimize code. It is maybe [...] 1/10 times that something works. Quite often Copilot insists on something that is not an improvement.
But that is not so bad, there are improvements in the end. [...] This reduces the time RevisionDataGridView.Add() runs from about 1000-1400 ms to 400-600ms}''. \faLightbulbO~When considering our findings for both \cg and Copilot, it is clear the need for strong judgments skill on the developers side in order to discriminate between good and bad recommendations from the LLMs. While this may seem more a lessons to learn for developers, it highlights the importance of studying such phenomenon to understand, for example, whether the usage of these AI-based recommenders may be counterproductive for novice developers, hindering their learning and resulting in suboptimal implementation choices.

When it comes to test generation, Copilot has been successfully used in several commits, issues, and PRs, see \eg: ``\emph{Quite cool, there are several nice and logical tests}''\footnote{https://github.com/tjarkvandemerwe/tidyprompt/pull/71}. \faLightbulbO~The complementarity between LLM-generated tests and more classic search-based generation techniques, such as Evosuite \citep{Fraser:fse2011}, represents an interesting avenue for research, especially to understand whether LLMs should be considered the new state-of-the-art tools for this problem, and whether its combination with search-based techniques would be valuable \citep{10.1145/3671016.3674813}.


\begin{figure*}
	\centering
	\includegraphics[width=1\linewidth]{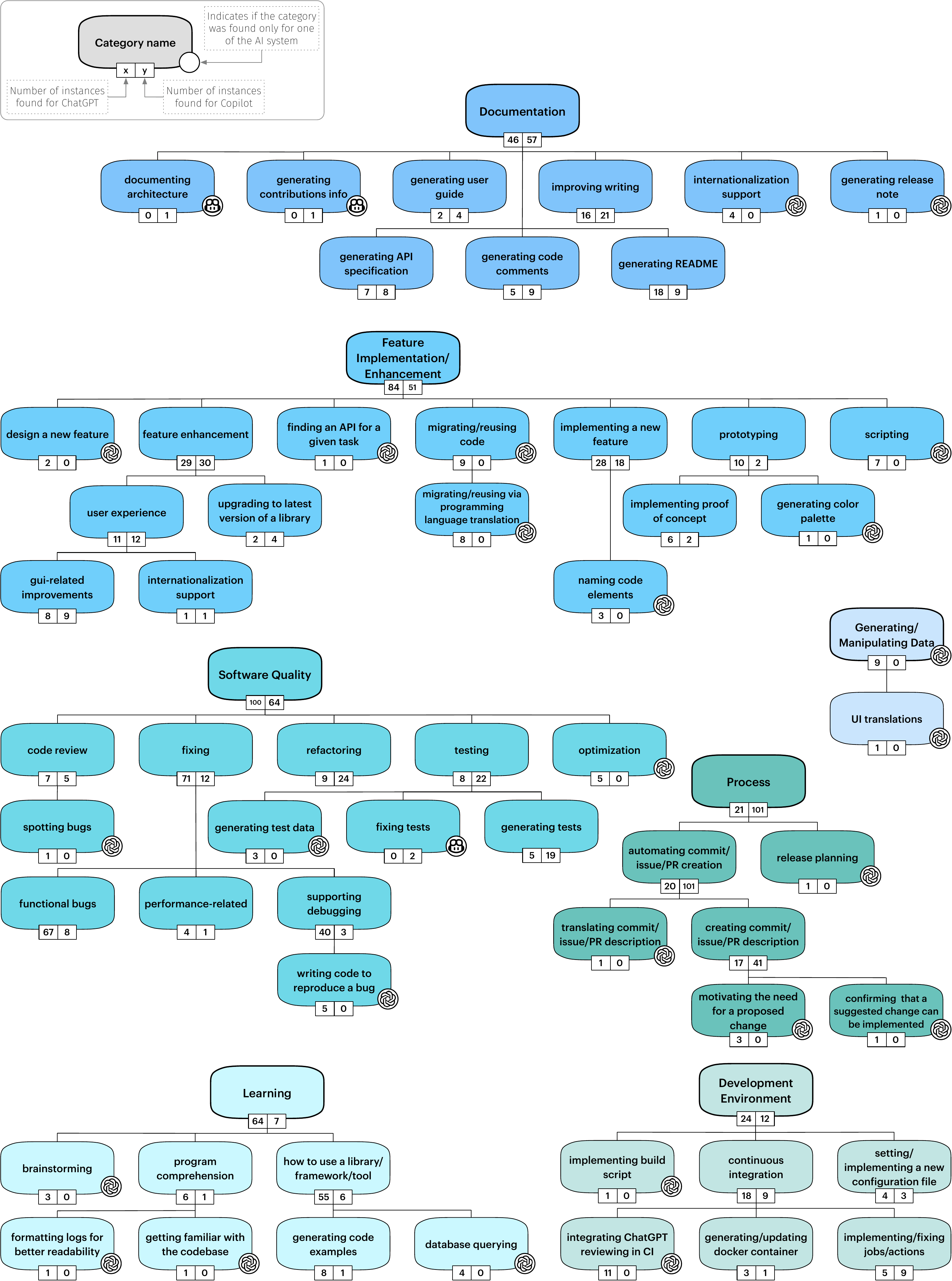}
	\caption{\rev{Taxonomy of tasks for which developers self-admitted the use of \cg and/or \cp}}
	\label{fig:taxonomyMerge}
\end{figure*}

\definecolor{lightlightgrey}{RGB}{233, 236, 239}

\begin{table*}
\centering
\caption{Summary of implications for practitioners derived from our study}
\label{tab:implicationsPractitioners}
{\footnotesize
\resizebox{1\textwidth}{!}{%
\begin{tabular}{p{14cm}}

\bottomrule
\rule{0pt}{4ex} \cellcolor{black} \textcolor{white}{\large \faLaptop \hspace{2 mm}  \bf \small Insights for Practitioners}  \\[2ex] \hline
\rule{0pt}{3ex} \cellcolor{lightlightgrey} \textbf{\small Contributions including AI-generated content}
\vspace{0.04cm}\\[1ex]

\begin{itemize}[leftmargin=0.5cm]
\setlength\itemsep{0.5em}
\item Define guidelines for projects' contributions including AI-generated code, \eg a project may decide to only welcome AI-generated code from users that are confident in assessing the correctness of the contributed code. \includegraphics[scale=0.025]{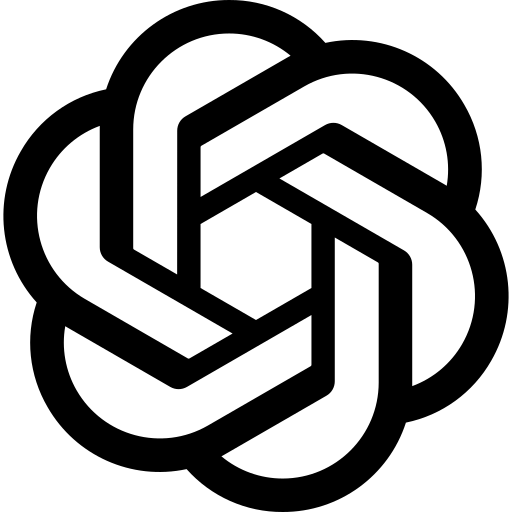}  \includegraphics[scale=0.021]{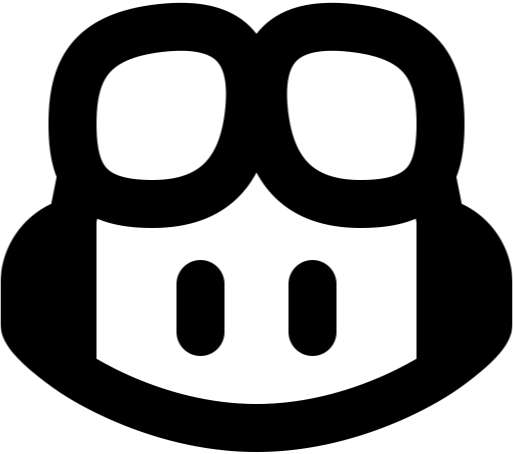}  \includegraphics[scale=0.025]{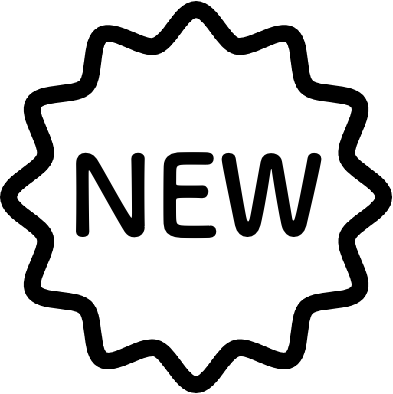} \includegraphics[scale=0.025]{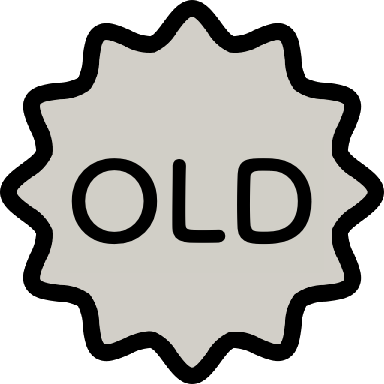} 

\item Adapt the code review process, \eg relying less on (semi-) automated code quality checks when AI- generated contributions come from users having little or no programming expertise. \includegraphics[scale=0.025]{img/chatgpt.png}  \includegraphics[scale=0.025]{img/new_.png} 

\item Clear risk related to the ownership and understanding of code contributed via \cg, especially when it is used to contribute with a complete feature: The (human) contributor is not always able to explain or advocate for the submitted code. \includegraphics[scale=0.025]{img/chatgpt.png} \includegraphics[scale=0.021]{img/copilot.png} \includegraphics[scale=0.025]{img/new_.png} \includegraphics[scale=0.025]{img/old.png} 

\item As with any AI-based solution, the usage of \cg for software-related tasks may result in artificial hallucinations: AI responses that look plausible to the user can be clearly wrong. The hard skills of developers remain essential in the era of AI-assisted coding.  \includegraphics[scale=0.025]{img/chatgpt.png} \includegraphics[scale=0.021]{img/copilot.png} \includegraphics[scale=0.025]{img/new_.png} \includegraphics[scale=0.025]{img/old.png}

\item Increasing importance of prompt engineering skills software developers must acquire to effectively exploit LLMs in their daily activities. \includegraphics[scale=0.025]{img/chatgpt.png} \includegraphics[scale=0.021]{img/copilot.png} \includegraphics[scale=0.025]{img/new_.png}

\end{itemize}\vspace{0.01cm}\\\hline


\rule{0pt}{3ex}  \cellcolor{lightlightgrey} \textbf{\small Automation possibilities offered by the AI}
\vspace{0.04cm}\\[1ex]

\vspace{0.04cm}LLMs can be leveraged to support very complex tasks, for which their usage has been only partially documented/experimented in the literature. These include:\vspace{0.1cm}

\begin{itemize}[leftmargin=0.5cm]
\setlength\itemsep{0.5em}
\item Prototyping the complete first version of a project, providing a substantial jumpstart in software development.  \includegraphics[scale=0.025]{img/chatgpt.png} \includegraphics[scale=0.021]{img/copilot.png} \includegraphics[scale=0.025]{img/old.png} \includegraphics[scale=0.025]{img/new_.png}

\item TDD collaboration, where the developer is mostly in charge of writing tests and delegating to LLM the code writing task.  \includegraphics[scale=0.025]{img/chatgpt.png} \includegraphics[scale=0.025]{img/old.png} 

\item Translating source code across different programming languages, thus improving code reusability.  \includegraphics[scale=0.025]{img/chatgpt.png} \includegraphics[scale=0.025]{img/new_.png} \includegraphics[scale=0.025]{img/old.png} 

\item Release planning, suggesting ideas on how to improve a software project based on what was observed in the wild.  \includegraphics[scale=0.025]{img/chatgpt.png} \includegraphics[scale=0.025]{img/new_.png} \includegraphics[scale=0.025]{img/old.png} 

\item Data generation, \eg augmenting UI-related strings handling dialogs with the user, generating test data.  \includegraphics[scale=0.025]{img/chatgpt.png} \includegraphics[scale=0.025]{img/new_.png} \includegraphics[scale=0.025]{img/old.png} 

\item Debugging, from several different perspectives, including helping in locating the bug as well as in reproducing it.  \includegraphics[scale=0.021]{img/copilot.png} \includegraphics[scale=0.025]{img/chatgpt.png} \includegraphics[scale=0.025]{img/new_.png} \includegraphics[scale=0.025]{img/old.png}

\item Brainstorming to steer towards the best implementation solution. \includegraphics[scale=0.025]{img/chatgpt.png} \includegraphics[scale=0.025]{img/new_.png}

\item AI-based review aimed at providing a first quick feedback to the contributor. \includegraphics[scale=0.025]{img/chatgpt.png} \includegraphics[scale=0.021]{img/copilot.png} \includegraphics[scale=0.025]{img/new_.png} \includegraphics[scale=0.025]{img/old.png}

\end{itemize}\vspace{0.01cm}\\\hline


\rule{0pt}{3ex} \cellcolor{lightlightgrey} \textbf{\small Software-related tasks involving natural language}
\vspace{0.04cm}\\[1ex]

\vspace{0.04cm} Due to their extensive training on natural language artifacts, \cg and Copilot are well-suited to support software-related tasks strongly characterized by natural language, such as the generation of software documentation.  \includegraphics[scale=0.025]{img/chatgpt.png} \includegraphics[scale=0.021]{img/copilot.png} \includegraphics[scale=0.025]{img/new_.png} \includegraphics[scale=0.025]{img/old.png} 
\vspace{0.4cm}\\\hline


\rule{0pt}{3ex} \cellcolor{lightlightgrey} \textbf{\small Risks related to sensible/private information}
\vspace{0.04cm}\\[1ex]

\vspace{0.04cm}Some of the tasks automated via LLMs, \eg code review, require to pass them sensible information, such as the code base itself, which may not be acceptable in industrial environments. Practitioners must carefully consider the trade-off of using a publicly available LLM \emph{vs} training a local LLM.  \includegraphics[scale=0.025]{img/chatgpt.png} \includegraphics[scale=0.025]{img/old.png} \vspace{0.4cm}\\\hline


\rule{0pt}{3ex} \cellcolor{lightlightgrey} \textbf{\small Unsuitability of AI for tasks dealing with recent technologies or requiring large contextual information}
\vspace{0.04cm}\\[1ex]

\begin{itemize}[leftmargin=0.5cm]
\setlength\itemsep{0.5em}
\item LLMs may not be suitable for tasks requiring up-to-date technology appeared after their last retraining. LLMs leveraging up-to-date knowledge available in the wild may obtain better results. \includegraphics[scale=0.025]{img/chatgpt.png} \includegraphics[scale=0.025]{img/new_.png} \includegraphics[scale=0.025]{img/old.png} 
\vspace{0.2cm}

\item LLMs may not be suitable for tasks requiring large contextual information (\eg knowledge of the whole code base), since their input length is limited. \includegraphics[scale=0.025]{img/chatgpt.png} \includegraphics[scale=0.025]{img/old.png}

\end{itemize}\\\hline

\end{tabular}

}
}

\end{table*}


\begin{table*}
\centering
\caption{Summary of implications for researchers derived from our study}
\label{tab:implicationsResearchers}
{\footnotesize
\resizebox{1\textwidth}{!}{%
\begin{tabular}{p{14cm}}

\rule{0pt}{4ex} \cellcolor{black} \textcolor{white}{\large \faLightbulbO \hspace{2 mm}  \bf \small Insights for Researchers}  \\[2ex] \hline
\rule{0pt}{3ex} \cellcolor{lightlightgrey} \textbf{\small Implications for the design of empirical studies}
\vspace{0.04cm}\\[1ex]

\begin{itemize}[leftmargin=0.5cm]
\setlength\itemsep{0.5em}
\item Empirical investigations studying OSS contributors  may or may not consider representative developers that only submitted AI-generated code.  \includegraphics[scale=0.025]{img/chatgpt.png} \includegraphics[scale=0.025]{img/new_.png} \includegraphics[scale=0.025]{img/old.png} 

\item \cg and Copilot must be considered as a baseline in works proposing novel recommenders for tasks where it was found to be useful. However, as the dataset on which these LLMs have been trained are not publicly available, it is hard to make a fair comparison ensuring the lack of overlap between training and test set. A possible solution is to use recent data points as test set, since those are unlikely to have been seen by the LLMs. \includegraphics[scale=0.025]{img/chatgpt.png} \includegraphics[scale=0.021]{img/copilot.png} \includegraphics[scale=0.025]{img/new_.png} \includegraphics[scale=0.025]{img/old.png} 

\item For researchers mining code from software repositories, it may be difficult to clearly discriminate the automatically generated from the manually written code, especially when the LLM is integrated in the IDE as in the case of Copilot. \includegraphics[scale=0.021]{img/copilot.png} \includegraphics[scale=0.025]{img/new_.png} 
\end{itemize}\vspace{0.01cm}\\\hline


\rule{0pt}{3ex} \cellcolor{lightlightgrey} \textbf{\small Studying and enhancing AI-aided development processes}
\vspace{0.04cm}\\[1ex]

\vspace{0.04cm}Practitioners are already leveraging LLMs for a variety of tasks. Nevertheless, it may be useful to (empirically) devise AI-enabled development processes, with suitable guidelines. These include using LLMs:
\vspace{0.1cm}

\begin{itemize}[leftmargin=0.5cm]
\setlength\itemsep{0.4em}
\item In TDD, with the developer being mostly in charge of writing tests and delegating to the LLM the production code.  \includegraphics[scale=0.025]{img/chatgpt.png} \includegraphics[scale=0.025]{img/old.png} 
\item To support program comprehension, especially when newcomers onboard a project and must become familiar with its code base.  \includegraphics[scale=0.025]{img/chatgpt.png} \includegraphics[scale=0.021]{img/copilot.png} \includegraphics[scale=0.025]{img/new_.png} \includegraphics[scale=0.025]{img/old.png} 
\item To generate tests.  \includegraphics[scale=0.025]{img/chatgpt.png} \includegraphics[scale=0.021]{img/copilot.png} \includegraphics[scale=0.025]{img/new_.png}  \includegraphics[scale=0.025]{img/old.png} 
\item To automate code review.  \includegraphics[scale=0.025]{img/chatgpt.png} \includegraphics[scale=0.021]{img/copilot.png} \includegraphics[scale=0.025]{img/new_.png} \includegraphics[scale=0.025]{img/old.png} 

\item To generate data possibly featuring unwanted discriminatory or offensive text.  \includegraphics[scale=0.025]{img/chatgpt.png} \includegraphics[scale=0.025]{img/new_.png} 

\end{itemize}\\\\

Moreover, developers seem to prefer in-IDE integration especially when it comes to tasks that require:
\begin{itemize}[leftmargin=0.5cm]
	\setlength\itemsep{0.4em}
\item Understand, refactor, complete very specific code or other artifacts.  \includegraphics[scale=0.025]{img/chatgpt.png} \includegraphics[scale=0.021]{img/copilot.png} \includegraphics[scale=0.025]{img/new_.png} \includegraphics[scale=0.025]{img/old.png} 
\item Avoid exposing internal artifacts to the outside, \eg using Retrieval Augmentation Generation or similar approaches.  \includegraphics[scale=0.025]{img/chatgpt.png} \includegraphics[scale=0.025]{img/old.png} 
\item Being highly repetitive and possibly benefiting of further integration with other used tools, such as the issue tracker (\eg generating PR descriptions). \includegraphics[scale=0.021]{img/copilot.png} \includegraphics[scale=0.025]{img/new_.png}
\end{itemize}\vspace{0.01cm}\\

Finally, it is important to carefully consider potential drawbacks of using AI-based assistants, for example in terms of code ownership (\ie are developers able to argue for their implementation choices?), knowledge retaining (\eg is it counterproductive for newcomers to use AI assistants for their first contributions in a novel code base?), and cognitive effort in assessing the AI recommendations. \includegraphics[scale=0.025]{img/chatgpt.png} \includegraphics[scale=0.021]{img/copilot.png} \includegraphics[scale=0.025]{img/new_.png}\\\\\hline

\rule{0pt}{3ex} \cellcolor{lightlightgrey} \textbf{\small Questioning the suitability of existing recommenders for AI-generated code}
\vspace{0.04cm}\\[1ex]

\vspace{0.04cm}The effectiveness of recommender systems for software engineers proposed in the literature may need to be reassessed on AI-generated code, since the latter may have characteristics different from those of human-written code. \includegraphics[scale=0.025]{img/chatgpt.png} \includegraphics[scale=0.025]{img/new_.png} \includegraphics[scale=0.025]{img/old.png} \vspace{0.04cm}\\\hline

\end{tabular}
}
}

\end{table*}

\section{Study Implications}
\label{sec:implications}

\rev{\figref{fig:taxonomyMerge} shows a merged taxonomy reporting all tasks for which developers self-admitted the use of \cg and/or \cp. Based on it, on our previous findings reported in \citep{tufano:msr2024}, and on the discussed qualitative examples,} \tabref{tab:implicationsPractitioners} and \tabref{tab:implicationsResearchers} summarize the implications our study has for practitioners and researchers, respectively. The icons reported in the tables indicate whether the insight: (i) has been derived from the manual analysis of \cg instances (\includegraphics[scale=0.025]{img/chatgpt.png}) and/or of Copilot instances (\includegraphics[scale=0.021]{img/copilot.png}); (ii) was already part of our findings in the MSR'24 paper (\includegraphics[scale=0.025]{img/old.png}); and (iii) has also been found in this study (\includegraphics[scale=0.025]{img/new_.png}), either confirming a previous insight (thus appearing together with the \includegraphics[scale=0.025]{img/old.png} icon) or being reported for the first time. 
While all insights can be directly read from the tables, we briefly discuss (i) the insights from our MSR'24 paper that have not been confirmed from the new sample; and (ii) the insights found for the first time in the new sample.

\subsection{Insights for Practitioners}

\textbf{Previous insights not confirmed.} There are three past insights \citep{tufano:msr2024} not confirmed in our study. The first, related to the usage of \cg in a TDD fashion, with developers writing tests and the LLM implementing the production code, was the result of a single instance we found in our previous work. However, while we could not find it in the new sample, this is nowadays a consolidated way of using LLMs for code generation \citep{Fakhoury24,10.1145/3691620.3695527}. For this reason, it is likely that with a larger sample inspected we could still find such a type of usage scenario.

The other two unconfirmed insights deserve more discussion. The first relates to risks concerning sensible/private information that must be shared to use the LLMs. Indeed, in our previous work we found instances discussing the fact that some tasks automated via LLMs, \eg code review, require to pass them sensible information, such as the code base itself, which may not be acceptable in some companies. The insight for practitioners was to carefully
consider the trade-off of using LLMs accessible through APIs versus a local deployment of (open) LLMs, which may require the availability of non-trivial computational and energy resources.
 In the meanwhile, features allowing the users of LLM-based assistants to opt out from the usage of the shared code/data for further training the models have been introduced, thus partially smoothing such a concern. Still, some companies may want to train and deploy local LLMs because, for example, they operate in niche scenarios for which commercial LLMs do not provide support, such as uncommon or even proprietary programming languages for which litter or no training data has been seen).

The second relates instead to the following observation: ``\emph{LLMs may not be suitable for tasks requiring large contextual information (\eg knowledge of the whole code base), since their input length is limited}''. Such an observation became, after two years, obsolete. Indeed, when we wrote the paper, the maximum input length of \cg was 8,192 tokens. Nowadays, GPT-4.1 accepts 1M tokens as input, allowing it to ingest even entire software repositories.

\textbf{Novel insights.} Some insights only emerged over the new sample, reflecting recent evolution of LLM-based assistants. The first insight relates to the need for adapting the code review process to rely less on automated quality checks when AI-generated contributions come from users having little expertise, either in terms of programming or project knowledge. Such a novel insight is likely due to the fact that LLMs are becoming better and better in confidently recommending code which looks correct, but which may hide difficult to catch bugs and/or suboptimal choices. 

The second insight concerns a novel automation possibility we observed for \cg, namely its usage as a brainstorming companion, which can help to steer towards the best implementation solution.

\subsection{Insights for Researchers}

\textbf{Previous insights not confirmed.} 
The unconfirmed insights for researchers from the previous taxonomy completely match those for practitioners, \ie the use of LLM for TDD, and avoiding exposing internal artifacts to the outside.


\textbf{Novel insights.}
While our study showed that in some cases developers self-admit the usage of LLMs in code writing, tools such as Copilot which continuously intervene at implementation time make it difficult for researchers to discriminate between human-written and LLM-written (either completely or partially) code. This hinders studies comparing the characteristics of the resulting code, necessary for a better understanding of pros and cons of AI-assisted development. This finding encourages, therefore, to conduct research in such a direction.

A second insight concerns the possible generation of unwanted, discriminatory, or offensive text from LLMs, such as when they are used to generate interaction content such as dialogues with users. Besides the more classic forms of testing, this sort of misbehaviors should also be caught at testing time. It is worthwhile noting how some LLMs, such as Meta Llama \citep{grattafiori2024llama3herdmodels}, have established terms of use forbidding the use of LLMs for the generation of offensive text.

Last but not least, there is an emerging preference for LLMs integrated with software development and maintenance tools (\ie IDE or issue trackers) which is visible from the very frequent usage of Copilot as support for generating issues, PRs, as well as for bug-fixing. In the long run, this may polarize the developers' preferences when it comes to their favorite LLM-based assistant.    
\section{Threats to Validity} \label{sec:threats}

Threats to \emph{construct validity} concern the relationship between the theory and observation. In principle, studying the purpose of the use of \cg and \cp  in software development by mining software repositories has an intrinsic limitation. This is because we observe only cases where developers mention \cg or \cp explicitly in a commit message, issue, or PR description. There could be other changes in which developers silently leveraged LLM-based assistants. However, our goal is not to understand the use of LLM in open-source development, which would have required not only a mining-based study, but, also, other research instruments, such as AI-generated code recognition, or developers' surveys/interviews. Indeed, we are interested to focus our analysis on all circumstances in which such an LLM usage was admitted, either because developers wanted to report their (positive or negative) experience, because they wanted to leave a trace, or because the generated artifact also contained a (generated) text acknowledging the use of the LLM-based assistant.

As explained in \secref{sec:design}, we analyzed the textual content of commit messages, issues, and PRs, as they could be queried by GitHub. However, there may be other places where \cg could have been mentioned, \eg code comments. 

A further threat is due to our interpretation of \cg purposes of usage, by reading and labeling commits and developers' discussions. This classification could have been affected by subjectiveness and imprecision. As explained in \secref{sec:design}, we mitigated this threat by having two annotators labeling each instance independently, and, after that, having a cooperative conflicts' resolution process. 

Threats to \emph{internal validity} concern confounding factors internal to our study that could affect our results. During the manual analysis, we explicitly excluded cases in which the contribution of the LLM-based assistants to a given development activity was unclear. Also, we used multiple labels in all cases for which an LLM-based assistant was used for multiple purposes.
There are many aspects of the LLM-based assistants usage we could not properly capture based on our observation sources. First of all, we could not know the exact \cg version, nor the LLM used by \cp (when \cp is used with its chat, it is possible to choose among multiple engines, such as different versions of GPT \citep{gpt4}, or Claude \citep{Anthropic_Claude3}). On the same line, we did not know, for \cg and for the \cp chat, the prompt being used by the developers, not other settings, such as the temperature.

Threats to \emph{external validity} concern the generalizability of our findings. First, this study is only limited to \cg and \cp. While these are among the most popular LLM-based assistants used by developers---the latter because of its IDE integration, the former because of its broad popularity---they are not the only ones. Also, the observed findings limit to open-source projects hosted on GitHub only.

\section{Related Work} \label{sec:related}

In this section, we start (\secref{sec:rel_software}) by highlighting the multifaceted capabilities of LLMs in addressing a broad spectrum of complex tasks, focusing on studies relying on state-of-the-art AI-based recommenders for software-related tasks such as ChatGPT~\citep{chatgpt} and GitHub Copilot~\citep{copilot}. After that, we review (\secref{sec:rel_admit}) the relevant literature concerning whether and to what extent developers admit the adoption of LLMs for software development tasks when contributing to open-source projects. 

\subsection{LLMs for Software-related Tasks}
\label{sec:rel_software}

Several researchers have analyzed the potential of LLMs to assist developers across different stages of the software development. \rev{In the following we will discuss only the works focusing on \cg and \cp, while a comprehensive analysis can be found in systematic literature reviews (SLRs) on the topic~\citep{fan2023large, HouZLYWLLLGW24}.} Specifically, the survey from \cite{fan2023large} points out that requirement engineering and software design are not comprehensively investigated by the research community, while code completion and generation is the area that has been most thoroughly explored. Last but not least, the authors highlight that currently the research community is focusing on evaluating the LLMs produced code. \cite{HouZLYWLLLGW24}, instead, conducted a SLR on LLM4SE revealing that LLMs are most widely used for code generation and program repair. 

\rev{Recent works also looked at differences between Agentic-PRs (\ie PRs created using an AI agent) and Human-PRs (\ie PRs created without the support of an AI agent). A representative example of this line of research is the study by \cite{watanabe2025useagenticcodingempirical}, that focused on one specific AI agent, namely Claude Code. Besides quantitatively comparing the two types of PRs, this work relates to ours since the authors also classified the goal of the PRs into one of ten possible categories (\eg fix, feature, refactor, style, performance, \etc). Our taxonomy, besides being related to different AI tools (\ie \cg and \cp): (i) has been defined without any predefined categories (tasks) in mind, emerging from our manual analysis; (ii) is the result of looking not only to PRs, but also to commits and issues; and (iii) relates to software artifacts spanning a much longer time period (18 \emph{vs} 2 months). Still, it is interesting to see that the categories found by \cite{watanabe2025useagenticcodingempirical} are covered in our taxonomy, while we highlighted several tasks not documented in their work. For example, \emph{code review} automation, the \emph{generation/manipulation of data}, and the usage of LLMs as a \emph{learning} mechanism was not documented by \cite{watanabe2025useagenticcodingempirical}.}

\rev{Finally, \cite{abs-2406-07765} surveyed 481 developers to collect their usage of AI across five broad categories of tasks: implementing new features, writing tests, bug triaging, refactoring, and writing natural-language artifacts. While the granularity of the obtained tasks substantially differ from ours making difficult a direct comparison, all these broad categories are represented in our taxonomy.}

\textbf{\cg for Software-related Tasks}

\cite{tian2023chatgpt} evaluated the effectiveness of using \cg for code generation and code summarization comparing its performance against state-of-the-art techniques. Their analysis reveal that \cg struggles in generating code for unfamiliar problems, confirmed in our study, and is sometimes less accurate in explaining the intention of a given code. A recent study by \cite{gu2025effectiveness} investigates the ability of several LLMs, including \cg and CodeLlama, in generating domain-specific code, focusing on Web development in Go and game development in C$++$. By comparing their performance on general-purpose and domain-specific tasks, the authors point out that LLMs tend to perform sub-optimally in domain-specific scenarios, although incorporating API knowledge into the prompts can enhance the performance.

Several researchers investigated the usage of \cg for automated program repair (APR). \cite{cao2025study} focus on program repair within deep learning applications, which are notably challenging to debug and test. Specifically, the authors evaluate \cg's ability to detect faulty programs, localize the faults, and perform automatic repairs. They also highlight that variations in prompts may significantly impact \cg's effectiveness in handling these tasks. \cite {sobania2023analysis}, instead, investigate the capability of \cg to fix bugs using the QuixBugs benchmark. Their results show that \cg has similar performance when compared to both traditional and deep learning-based APR approaches. From a different perspective, \cite{fan2023automated} investigate to what extent APR techniques can be used to fix erroneous code generated by LLMs. Their findings reveal that LLM-generated code often contains common errors also found in human-written code, indicating that APR techniques can effectively be applied to AI-generated code as well. These studies reinforce the potential of \cg in the context of APR, \ie a task that, according to our taxonomy, developers frequently rely on it for.

\cite{guo2024exploring} investigate the capability of \cg in performing code review tasks, \ie automated code refinement based on given code reviews and compare its performance with CodeReviewer highlighting that \cg outperforms CodeReviewer in code refinement tasks. \cite{liu2025exploring}, instead, investigate the feasibility to rely on LLMs, \cg and Gemini, for automated software refactoring, focusing on the identification of refactoring opportunities and the recommendation of refactoring strategies. Unfortunately, the results of the study reveal that the automated software refactoring is still an open problem, although the performance improves guiding the LLM in the original prompt, such as by narrowing the search space. 

Last but not least, another area of interest for the research community is test generation. Specifically, \cite{yuan2024evaluating}, evaluate the \cg's ability to generate unit tests in terms of correctness, completeness, readability and usability. Their findings reveal that, even if \cg-generated test often face correctness issues, those that pass closely resemble human-written tests in terms of coverage and readability. 

\textbf{GitHub Copilot for Software-related Tasks}

Numerous studies have explored the use of GitHub Copilot \citep{copilot}, covering various aspects such as its impact on developers \citep{imai2022github, vaithilingam2022expectation, xiao2024generative, ziegler:maps2022}, assessments of its robustness \citep{mastropaolo2023robustness, yetistiren2022assessing, wong2022exploring}, empirical evaluations of the correctness of its suggestions \citep{nguyen2022empirical, yetistiren2022assessing}, and investigations into its security implications \citep{fu2025security, pearce2021empirical, sobania2021choose, asare2022github}.

\cite{imai2022github} investigate the extent to which Copilot can serve as a viable alternative to a human pair programmer. In their study, 21 participants completed coding tasks under three conditions: (i)  pair-programming with Copilot; (ii) human pair-programming as a driver; and (iii); human pair-programming as a navigator. The results showed that using Copilot led to higher productivity but resulted in lower code quality. \cite{vaithilingam2022expectation}, on the other hand, found that Copilot does not significantly enhance task completion time or success rate. Nonetheless, developers appreciated its assistance, noting that its suggestions serve as useful starting points, reducing the need for external searches. A similar result has been obtained in the study by \cite{ziegler:maps2022}. Specifically, by examining the relationship between Copilot usage and developers' productivity, the authors state that the acceptance rate of suggested solutions can act as a reliable proxy for perceived productivity. Last but not least, \cite{xiao2024generative} empirically analyzed the adoption of GitHub Copilot for (partially) generating PR descriptions and its effects on the development workflow. The findings of the study suggest that Copilot-assisted PRs tend to be reviewed more quickly and are more likely to be merged, however the generated descriptions are frequently adjusted with manual edits. 

\cite{mastropaolo2023robustness} analyzed the robustness of GitHub Copilot in automatically generating Java methods. By supplying Copilot with several semantically equivalent descriptions of the methods to be generated, they point out that the wording of the prompt can significantly influence the output. 

\cite{yetistiren2022assessing} aimed to evaluate the quality of code generated by GitHub Copilot, focusing on factors such as validity, correctness, and efficiency. Their findings highlighted that Copilot generated syntactically valid code with a high success rate of 91.5\%. However, further analysis showed that only 28.7\% of the completed tasks were fully correct, 51.2\% were partially correct, and 20.1\% were incorrect. 

Moving into the security area, \cite{fu2025security} used regular expressions to detect code snippets generated by GitHub Copilot, CodeWhisperer and Codeium in Python and Javascript GitHub projects to assess the security of automatically generated code. Their analysis shows that approximately 30\% of these snippets contain at least one security vulnerability, regardless of the programming language. Although the overall incidence is similar between Python and JavaScript, the types of identified  CWEs differ. Indeed, Python is more prone to vulnerabilities related to data processing and system calls, whereas JavaScript is mainly affected by issues stemming from dynamic code generation and security flaws in the web development. \cite{pearce2021empirical} found that 40\% of GitHub Copilot's code recommendations contained security vulnerabilities. This result has been confirmed by \cite{asare2023github} revealing that Copilot introduced vulnerabilities in about 33\% of cases when generating code in contexts in which human developers had introduced vulnerabilities in the past. 

All the previously mentioned studies evaluate the feasibility of relying on LLMs, namely \cg and GitHub Copilot, to support developers during different stages of the software development process. Our study, instead, tries to characterize what are the stages in which real developers seek the assistance of LLMs by looking at their ``admission''. 

\subsection{Empirical Studies on how developers ``admit'' the use of AI-based recommenders}
\label{sec:rel_admit}

\cite{xiao2024devgpt} were the first curating a dataset, \textsc{DevGPT}, accounting for conversations between developers and \cg collected from two sources, GitHub and Hacker News. In curating the dataset the authors relied on a feature released by OpenAI that allows the sharing of \cg interactions, including prompts with \cg's responses and their related code snippets, through dedicated links. Upon its release, the dataset has been used by several researchers to shed light on how \cg is used by software developers by looking directly at their interactions. Specifically, \cite{sagdic2024taxonomy} proposed a taxonomy comprising 17 topics, organized into seven categories, based on developers' interactions with \cg across issues, PRS and GitHub discussions. Analyzing the frequency for each category, the authors found that over half of the examined prompts relate to programming guidance, confirming \cg's pivotal role in supporting code generation and problem-solving. 

\cite{siddiq2024quality}, focusing only on Java and Python projects, revealed that (i) \cg is used as a support for every stage of the software development, \eg code generation, refactoring, testing, debugging and documentation, and (ii) developers mostly use it to learn about new frameworks as a replacement for official documentation and Q\&A websites. \cite{champa2024chatgpt}, instead, other than identifying the development tasks within which developers rely on \cg, tried also to determine the extent to which the provided support is considered as ``beneficial'' from the developers' perspective. Quite surprisingly, documentation generations and code quality management are the least efficient ones. At the same time, interactions with \cg are regarded as effective for tasks related to software development management, optimization, and the implementation of new features. 

\cite{jin2024can} used the \textsc{DevGPT} dataset to understand the extent to which developers integrate \cg-generated code in production by qualitatively analyzing the merged Pull Requests (PRs), \ie comparing the automatically generated code with the code that is actually added in files modified within the PR. Confirming the findings by \cite{siddiq2024quality}, the authors highlighted that \cg is mostly used to demonstrate high-level concepts or providing examples in documentation, rather than to be used as production-ready code, \ie only in 16.8\% PRs the merged code is an ``exact replica'' of the one automatically generated. Similarly, \cite{grewal2024analyzing} looked at how \cg-generated code is integrated into real code confirming that \cg code is not ready to be merged while it requires modifications before being integrated. Furthermore, the authors conducted an evolutionary analysis to assess how extensively the integrated \cg-generated code is modified over time. Results show that the automatically generated code is rarely altered after integration, with most changes limited to minor functional adjustments, code restructuring, and improvements in naming. \cite{chouchen2024software}, instead, compared the characteristics of PRs containing at least one interaction with \cg to those that do not revealing that developers tend to use \cg in larger, more review-intensive PRs, primarily for explanation and code generation purposes.

From a different perspective, \cite{watanabe2024use} not only categorized the reasons why developers use \cg during code review, but also looked at how they react to the generated answers. Following the methodology by \cite{xiao2024devgpt}, they enriched the \textsc{DevGPT} dataset by extending the time interval. By manually inspecting 229 review comments across 205 PRs from 179 GitHub projects, the study revealed that developers use \cg for problem solving, just as often as they seek assistance for learning and for validating their opinions. Furthermore, the authors noted that developers do not always react positively to \cg's answers primarily because the answers fail to provide additional value. On a similar research trend, \cite{hao2024engaging} not only investigated the reactions of developers using \cg when sharing their conversations, but also analyzed the reactions from other developers to PR comments containing links to \cg interactions. Their findings differ somewhat from those by \cite{watanabe2024use}, primarily because their analysis was limited to data from the \textsc{DevGPT} dataset. Specifically, they noted that over half of the developers who used \cg expressed positive sentiment towards its answers/solutions, reporting the support as valuable. However,  it was also noted that developers involved in the PRs often did not react to comments containing \cg interactions. In a follow up study, \cite{hao2024empirical} conducted a more in-depth investigation into the dynamics of the sharing behavior, focusing not only on PRs but also considering the issues. Their findings reveal that developers ``acknowledge'' the use of \cg to support their role-specific contributions, \eg by clarifying that \cg was used as the basis for a solution, to come up with alternative approaches, or to reinforce their claims. 

Differently from previously mentioned studies, our work aims at creating a fine-grained taxonomy of \cg and GitHub Copilot usage tasks by mining GitHub for traces of ``self-admitted'' use of AI tools in commits, issues and PRs. Indeed, we do not look at shared links pointing to interactions with generative AI tools, while mine candidate instances searching for the word ``\cg'' and/or ``copilot''.

Last but not least, \cite{yu2024large} performed a large-scale mining study to characterize AI-generated code snippets, \ie accompanied by comments acknowledging the use of AI tools, as well as the GitHub projects disclosing such usage. Their findings indicate that (i) \cg and GitHub Copilot are the most commonly used AI tools for generating Python and Javascript code, (ii) projects containing AI-generated code tend to be small and maintained by small teams, (iii) the generated code is usually brief and relatively simple, and (iv) code comments rarely include detailed context, such as the original prompts.

\section{Conclusion}
\label{sec:conclusion}
Large Language Models (LLMs) based assistants, including Web-based ones such as \cg or IDE-integrated ones such as GitHub \cp, are radically changing the way practitioners are developing software. In this paper we have reported the results of an empirical study analyzing traces left by developers in commit messages, pull requests, and issues of open-source projects hosted on GitHub about the use of \cg and \cp to automate software development tasks. This study has extended our previous work \citep{tufano:msr2024} in which we analyzed \cg traces found on June 2023.

Specifically, we mined traces of \cg and \cp usages' admissions taken on January 2025 and manually coded \newlyInspectedChatGPT instances of commits, issues and pull requests from \cg and \newlyInspectedCopilot from \cp to obtain a statistically significant sample of \newlyValidChatGPT for \cg and a total set of \finalCopilot for \cp. 

On the one hand, results confirmed our previous findings about a very diverse type of LLM usage to automate tasks such as feature development/enhancement, software quality improvement, (re)documentation, software process support, or technology learning. On the other hand, we found that some challenges previously encountered have been greatly mitigated, \eg concerns about the LLMs' limited context window and risks related to seeding private content into LLMs. For \cp, we found a great support for process task automation, \eg commit and pull request generation, but also witnessed a limited variety of tasks, possibly because of its different usage scenario and because its IDE integration makes developers less prone to ``admit'' LLM usage.

As a side note, and has we have clarified throughout the paper, while on the one hand this paper shed the lights on LLM opportunities for software engineering automation, what we have observed may simply represent the ``tip of the iceberg'' as software repositories likely contain plenty of LLM-generated artifacts for which there is no explicit admission. Similarly, there may be several cases in which LLM failed and developers have not reported them.

As possible future directions, \rev{we plan to address the limited generalizability of our work by extending our analysis to further LLM-based assistants and to projects not hosted on GitHub (\eg commercial products).} It is also part of our work-in-progress to better understand, \eg by interviewing practitioners and by conducting ethnographic studies, what are the LLM admission ``good practices'' in different types of software organizations. Finally, we would like to explore better ways to support developers in documenting and tracking LLM-generated artifacts, so that such information is available during software evolution, as well as for transparency purpose, \eg as part of Software Bills of Materials (SBOMs).

\section*{\textbf{Fundings}}
Di Penta acknowledges the Italian ‘‘PRIN 2022’’ project TRex-SE: ‘‘Trustworthy Recommenders for Software Engineers’’, grant n. 2022LKJWHC.  Pepe is partially
funded by the PNRR DM 352/2022 Italian Grant for Ph.D. scholarships.
Tufano and Bavota have been funded by the European Research Council (ERC) under the European Union's Horizon 2020 research and innovation programme (grant agreement No. 851720).

\section*{\textbf{Data Availability Statement}}
Code and data used for conducting the study are publicly available in our replication package \citep{replication}.

\balance
\bibliographystyle{spbasic}
\bibliography{main}

%
%

\end{document}